\documentclass[12pt]{article}
\usepackage[margin=1in]{geometry}
\usepackage{amsmath}
\usepackage{amsfonts}
\usepackage{graphicx}
\usepackage{psfrag,epsf}
\usepackage{enumerate}
\usepackage{natbib}
\usepackage{url} 
\usepackage{prodint}
\usepackage{amsthm}
\usepackage{multirow}
\usepackage{booktabs}
\usepackage[colorlinks,citecolor=blue,urlcolor=blue]{hyperref}
\usepackage{algorithm2e}
\RestyleAlgo{ruled}
\LinesNumbered
\usepackage{longtable}
\usepackage{setspace}

\newtheorem{thm}{Theorem}

\newcommand{\blind}{1}
\newcommand{\GG}[1]{}

\begin{document}

\def\spacingset#1{\renewcommand{\baselinestretch}%
{#1}\small\normalsize} \spacingset{1}


\if1\blind
{
  \title{\bf Marginal Regression on Transient State Occupation Probabilities with Clustered Multistate Process Data}
  \author{Wenxian Zhou\hspace{.2cm}\\
    Department of Biostatistics and Health Data Science, Indiana University\\
    Giorgos Bakoyannis\hspace{.2cm}\\
    Department of Biostatistics and Health Data Science, Indiana University\\
    Ying Zhang\hspace{.2cm}\\
    Department of Biostatistics, University of Nebraska Medical Center\\
    Constantin T. Yiannoutsos\hspace{.2cm}\\
    Department of Biostatistics and Health Data Science, Indiana University\\
    }
  \maketitle
} \fi

\if0\blind
{
  \bigskip
  \bigskip
  \bigskip
  \begin{center}
    {\LARGE\bf Marginal Regression on Transient State Occupation Probabilities with Clustered Multistate Process Data}
\end{center}
  \medskip
} \fi

\bigskip
\spacingset{2} 
\begin{abstract}
Clustered multistate process data are commonly encountered in multicenter observational studies and clinical trials. A clinically important estimand with such data is the marginal probability of being in a particular transient state as a function of time. However, there is currently no method for nonparametric marginal regression analysis of these probabilities with clustered multistate process data. To address this problem, we propose a weighted functional generalized estimating equations approach which does not impose Markov assumptions or assumptions regarding the structure of the within-cluster dependence, and allows for informative cluster size (ICS). The asymptotic properties of the proposed estimators for the functional regression coefficients are rigorously established and a nonparametric hypothesis testing procedure for covariate effects is proposed. Simulation studies show that the proposed method performs well even with a small number of clusters, and that ignoring the within-cluster dependence and the ICS leads to invalid inferences. The proposed method is used to analyze data from a multicenter clinical trial on recurrent or metastatic squamous-cell carcinoma of the head and neck with a stratified randomization design.
\end{abstract}

\noindent%
\vfill

\section{Introduction}
\label{intro}
Disease evolution in many chronic illnesses is often characterized by multiple discrete health states \citep{putter2007tutorial}. For example, in oncology, disease evolution is frequently described by the states ``initial cancer state", ``tumor response" (i.e., tumor shrinkage according to the RECIST 1.0 \citep{RECIST} or RECIST 1.1 \citep{RECIST2} criteria), ``disease progression", and ``death". In such cases, the resulting event histories are known as multistate processes. A key estimand with multistate processes is the state occupation probability, which is defined as the probability of being in a particular state as a function of time. In the oncology trial context, the state occupation probability of tumor response \citep{temkin1978analysis, begg1982study, ellis2008analysis} is crucial for treatment efficacy evaluations since the tumor response, unlike the progression-free or overall survival, reflects a direct biological effect of treatment on the tumor \citep{Kaufman17}. In addition, the probability of being in the response state may be a more important outcome than the crude survival time since it reflects both quality and quantity of life \citep{Kaufman17}. Last but not least, a sustained tumor response is associated with a prolonged treatment-free interval \citep{Kaufman17}, which is related to lower costs, in addition to fewer side effects, for the patient.

The state occupation probabilities of transient states, such as the tumor response state, are non-monotonic functions of time. Therefore, nonparametric and semiparametric analyses of these probabilities cannot be performed using standard methods for survival or competing risks data \citep{bluhmki2018wild}, and require methods for multistate models. With right-censored and independent observations, nonparametric estimation of state occupation probabilities can be achieved using the Aalen--Johansen estimator \citep{aalen1978empirical}. \citet{datta2001validity} showed that the latter estimator is consistent for state occupation probabilities even when the multistate process of interest is non-Markovian. Methodology for simultaneous confidence bands and two-sample nonparametric tests have been proposed by \citet{bluhmki2018wild} and \citet{bakoyannis2020nonparametric}, respectively. The issue of nonparametric/semiparametric regression of state occupation probabilities with independent data has also been addressed in the literature \citep{fine2004temporal, andersen2007regression, andersen2008inference, azarang2017direct}. A limitation of the aforementioned methods is that they are not applicable to situations with cluster-correlated observations. Nevertheless, within-cluster dependence is ubiquitous in multicenter observational studies and clinical trials. 

For the cluster-correlated data setting under right censoring, \citet{bakoyannis2021nonparametric} proposed moment-based nonparametric estimators of the state occupation probabilities and nonparametric two-sample Kolmogorov--Smirnov-type tests. These tests are applicable to situations where every cluster in the sample contains observations from both groups under comparison (complete cluster structure). \citet{bakoyannis2022nonparametric} proposed nonparametric two-sample linear and $l_2$-norm-based tests for situations with or without complete cluster structure. The methods by \citet{bakoyannis2021nonparametric} and \citet{bakoyannis2022nonparametric} do not impose Markov assumptions or assumptions regarding the within-cluster dependence. Furthermore, they allow for informative cluster size (ICS), a situation where the number of observations within a cluster is associated with the outcomes in that cluster \citep{cong2007marginal, williamson2008modeling, pavlou2013examination, seaman2014methods}. Failure to account for this issue can lead to biased inferences, since larger clusters (e.g., clinics) contribute more observations in the study sample and, thus, they have a larger influence on the estimates. However, the methods by \citet{bakoyannis2021nonparametric} and \citet{bakoyannis2022nonparametric} do incorporate covariates. To the best of our knowledge, there is currently no method for nonparametric or semiparametric regression on state occupation probabilities with clustered multistate process data.

This research was motivated by the SPECTRUM trial, a multicenter two-arm clinical trial on recurrent or metastatic squamous-cell carcinoma of the head and neck \citep{vermorken2013cisplatin}. The treatments under comparison were chemotherapy combined with panitumumab and chemotherapy alone. In this trial, patient event history was characterized by the states “initial cancer state”, “tumor response” (per \citet{RECIST}), ``disease progression", and ``death”. There are several aspects of the SPECTRUM trial that require special attention in the data analysis. First, patients from the same clinic are expected to have correlated outcomes. Second, the cluster sizes are likely informative as tumor response is expected to be more frequent in larger clinics (which are typically better staffed and may provide better care; evidence for ICS in this trial was provided by a previous analysis presented in \citet{bakoyannis2022nonparametric}). Third, the SPECTRUM trial used a stratified randomization design, with the stratification variables being Eastern Cooperative Oncology Group (ECOG) performance status, prior treatment history, and site of primary tumor. It is well known that, not adjusting for the stratification variables can lead to statistical efficiency loss that results in lower power \citep{forsythe1987validity, pocock2002subgroup, white2005adjusting, zhang2008improving, kahan2012reporting, kahan2012improper} to statistically demonstrate treatment efficacy \citep{kahan2012reporting}. A previous analysis of the SPECTRUM data by \citet{bakoyannis2022nonparametric} addressed the lack of independence within clinics and the ICS issue, but did not adjust for the stratification variables. Of note, the latter analysis provided a statistically non-significant difference between the two treatments in terms of the state occupation probability of tumor response.

The goal of this work is to address the issue of marginal regression analysis of transient state occupation probabilities with right-censored multistate process data with cluster-correlated observations. To achieve this, we extend the temporal processes regression framework for independent observations \citep{fine2004temporal} to the cluster-correlated data setting, allowing also for ICS. Our estimation method relies on a weighted functional generalized estimating equations approach. The proposed methodology does not impose Markov assumptions or assumptions regarding the structure of the within-cluster dependence, and effectively addresses the complications in the SPECTRUM trial mentioned above. The asymptotic properties of the proposed method are rigorously established, and closed-form variance estimators are provided. Calculation of simultaneous confidence bands for the functional regression coefficients and p-values for the covariate effects are based on a wild bootstrap approach. Simulation studies provide numerical evidence that the proposed method works well even with a relatively small number of clusters and under ICS. The proposed methodology is used to analyze the data from the SPECTRUM trial. The goal of this analysis is to evaluate the effect of combining chemotherapy with panitumumab on the transient state occupation probability of the tumor response, while adjusting for the stratification variables and other potentially important baseline covariates. 

The rest of the paper is structured as follows. In Section~\ref{s:method}, we introduce the model and the proposed methodology, describe the asymptotic properties of the estimators, and provide a nonparametric hypothesis testing procedure for the covariate effects. In Section~\ref{s:sim}, we present a series of simulation studies to evaluate the finite sample performance of the proposed method and compare them with the previous methodology that ignores the within-cluster dependence and the ICS. Our methodology is applied to the SPECTRUM data for illustration in Section~\ref{s:data}. Finally, the paper concludes with a discussion in Section~\ref{s:discuss}.

\section{Methodology}
\label{s:method}

\subsection{Notation and Assumptions}

Consider a study with $n$ clusters (e.g., clinics) of observations from a multistate process $\{Y(t): t \in [0, \tau]\}$, where $\tau<\infty$ is the maximum follow-up time, with a finite set of states $\mathcal{S}=\{1, \ldots, k\}$ and a subset $\mathcal{T} \subset \mathcal{S}$ including possible absorbing states (e.g., death). Let $\{Y_{ij}(t): t \in [0, \tau]\}$ be the multistate process corresponding to the $j$th individual, with $j=1, \ldots, M_{i}$, in the $i$th cluster ($i=1, \ldots, n$). For the sake of generality, the cluster sizes $M_i$, $i=1, \ldots, n$, are considered to be random and informative, that is, the multistate processes are allowed to depend on the size of the corresponding cluster $M_i$. However, our proposed methodology is trivially applicable to simpler situations under non-informative or fixed cluster size. In this paper, we focus on the analysis of transient states $h \in \mathcal{T}^c$, which, as mentioned in the Introduction section, cannot be analyzed using standard methods for clustered survival or competing risks data \citep{bluhmki2018wild}. In contrast, the analysis of absorbing states $h \in \mathcal{T}$ can be based on standard methods for clustered survival and competing risks data. Let $\{\tilde{Y}_{ij, h}(t)=I[Y_{ij}(t)=h]: t \in [0, \tau]\}$, with $h \in \mathcal{T}^c$, $i=1, \ldots, n$, and $j=1, \ldots, M_{i}$, be the binary response processes indicating whether the $j$th individual from the $i$th cluster is at the transient state $h$ at time $t$. Now, the marginal state occupation probability for the state $h \in \mathcal{T}^c$ of scientific interest can be expressed as 
$$ 
P_{0,h}(t)=P[Y(t)=h]=E\{I[Y(t)=h]\}=E[\tilde{Y}_{h}(t)].
$$ 
In addition to the multistate and binary response processes, let $\boldsymbol{X}_{ij}(t)$ be a $(p+1)$-dimensional covariate vector, including the intercept term (constant $1$) and $p$ time-invariant and/or external time-dependent covariates, which are potentially associated with the state occupation probability of state $h$. Furthermore, denote by $S_{ij, h}(t)$ a time-dependent indicator variable with $S_{ij, h}(t)=1$ if the $j$th individual from the $i$th cluster has not reached to an absorbing state and is at risk for state $h \in \mathcal{T}^c$ at time $t$. Also, let $R_{ij, h}(t)$ be a missingness indicator, with $R_{ij, h}(t)=1$ if $(\tilde{Y}_{ij, h}(t), \boldsymbol{X}_{ij}(t), S_{ij, h}(t))$ is fully observed at time $t$, and $R_{ij, h}(t)=0$ otherwise (e.g., in the case of right censoring prior to time $t$). For $h\in\mathcal{T}^c$, the observed data are assumed to be independent and identically distributed (i.i.d.) copies of $\boldsymbol{O}_{i,h}=\{(\tilde{Y}_{ij, h}(t),\boldsymbol{X}_{ij}(t), S_{ij, h}(t)):R_{ij, h}(t)=1, j=1, \ldots, M_{i}, t\in[0,\tau]\}$, for $i=1,\ldots,n$ and $j\in\mathcal{T}^c$. Assuming the existence of the latent processes $\tilde{Y}_{ij, h}(\cdot),\boldsymbol{X}_{ij}(\cdot), S_{ij, h}(\cdot),R_{ij, h}(\cdot)$, for $j=M_i+1, \ldots, m_0$, where $m_0$ is an upper bound for the cluster sizes (see regularity condition C3 in Section 2.3), we can define $\boldsymbol{\tilde{O}}_{i,h}=\{(\tilde{Y}_{ij, h}(t),\boldsymbol{X}_{ij}(t), S_{ij, h}(t)):R_{ij, h}(t)=1, j=1, \ldots, m_0, t\in[0,\tau]\}$, for $i=1,\ldots,n$ and $h\in\mathcal{T}^c$. Then,  the i.i.d. assumption imposed in this article is implied if $(\boldsymbol{\tilde{O}}_{i,h},M_i)$ are identically distributed for $i=1,\ldots,n$, in addition to the independence assumption across clusters. The aforementioned latent processes do not contribute to our estimators but are assumed to exist for technical reasons, similarly to prior work on clustered data with random cluster sizes \citep[see e.g.][]{Cai00}. These latent processes can be seen as data of potential candidate individuals that could be included in the $i$th cluster (e.g., future patients that will attend the $i$th clinic). Independent and identically distributed observations assumptions across clusters are standard in the literature of statistical methods for clustered data with varying cluster sizes \citep[see e.g.][]{Cai00,Zhang11,liu2011positive,Zhou12,bakoyannis2021nonparametric,bakoyannis2022nonparametric,zhou2022}.

For simplicity of presentation, and without loss of generality, we focus on a single transient state $h$ of scientific interest. In this work, we impose the missing at random (MAR) assumption that the response $\tilde{Y}_{ij, h}(t)$ and missingness indicator $R_{ij, h}(t)$ are independent conditionally on $[\boldsymbol{X}_{ij}(t), S_{ij, h}(t)=1]$, that is
$$
P[\tilde{Y}_{h}(t)=1|R_{h}(t), \boldsymbol{X}(t), S_{h}(t)=1] = P[\tilde{Y}_{h}(t)=1|\boldsymbol{X}(t), S_{h}(t)=1],
$$
for $t\in[0,\tau]$. In addition, we assume that the mean of the response $\tilde{Y}_{ij, h}(t)$ at time $t$, conditionally on $\boldsymbol{X}(t)$ and $S_{h}(t)$ 
has the form
\begin{eqnarray}
E[\tilde{Y}_{h}(t)|\boldsymbol{X}(t), S_{h}(t)=1] = g_h^{-1}[\boldsymbol{\beta}_{0,h}^{T}(t)\boldsymbol{X}(t)], \quad t\in[0,\tau],
\label{eq1}
\end{eqnarray}
where the link function $g_{h}$ is a monotone, differentiable, and invertible function, and $\boldsymbol{\beta}_{0,h}(t)=(\beta_{0,h0}(t),\ldots,\beta_{0,hp}(t))^T$, $t\in[0,\tau]$, is the true functional regression coefficient vector which is an unspecified function of time. A choice for the link function $g_{h}$ is the complementary log-log link function $g_{h}(\theta)=\operatorname{cloglog}(\theta)=\log [-\log (1-\theta)]$, which provides a time-indexed marginal complementary log-log model. Another choice is the logit link function $g_{h}(\theta)=\operatorname{logit}(\theta)=\log (\theta/{1-\theta})$, which provides a
time-indexed marginal logistic model. Under the latter choice, $\boldsymbol{\beta}_{0,h}(t)$ possess a marginal log odds ratio interpretation.

\subsection{Estimation}

In this work, we propose a framework for marginal regression analysis of transient state occupation probabilities for clustered multistate process data. More precisely, we extend the framework for temporal process regression proposed by \citet{fine2004temporal} to account for within-cluster dependence and allow for ICS. For the estimation of the functional regression coefficients $\boldsymbol{\beta}_{0,h}(t)$ in \eqref{eq1} we propose the following weighted (by the inverse of the cluster size) functional generalized estimating equations approach. Under a working independence assumption, $\boldsymbol{\hat{\beta}}_{n,h}(t)$ is the root of the equation
\begin{eqnarray}
\boldsymbol{U}_h[\boldsymbol{\beta}_{h}(t),t]=\sum_{i=1}^{n}\frac{1}{M_i}\sum_{j=1}^{M_i} \boldsymbol{A}_{ij,h}[\boldsymbol{\beta}_{h}(t),t]=\mathbf{0}, \quad t\in[0,\tau],
\label{eq2}
\end{eqnarray}
and 
\begin{eqnarray*}
\boldsymbol{A}_{ij,h}[\boldsymbol{\beta}_{h}(t),t]=S_{ij, h}(t)R_{ij, h}(t)\boldsymbol{D}_{ij, h}^T[\boldsymbol{\beta}_{h}(t)]V_{ij, h}[\boldsymbol{\beta}_{h}(t),t]\{\tilde{Y}_{ij, h}(t)-g_h^{-1}[\boldsymbol{\beta}_{h}^{T}(t)\boldsymbol{X}_{ij}(t)]\},
\end{eqnarray*}
where $\boldsymbol{D}_{ij, h}[\boldsymbol{\beta}_{h}(t)]=d\{g_h^{-1}[\boldsymbol{\beta}_{h}^{T}(t)\boldsymbol{X}_{ij}(t)]\} / d [\boldsymbol{\beta}_{h}(t)]$ and $V_{ij, h}[\boldsymbol{\beta}_{h}(t),t]$ is a possibly random weight function. A natural choice for $V_{ij, h}[\boldsymbol{\beta}_{h}(t),t]$ is the inverse of the conditional variance of $\tilde{Y}_{ij, h}(t)$ given $\boldsymbol{X}_{ij}(t)$.  If the covariates are time-invariant or piecewise constant, then the estimator $\boldsymbol{\hat{\beta}}_{n,h}(t)$, $t \in [0, \tau]$, is also piecewise constant with jumps at $K_h$ time points where the observed data $\{(\tilde{Y}_{ij, h}(t),\boldsymbol{X}_{ij}(t), S_{ij, h}(t)):R_{ij, h}(t)=1\}$ and $R_{ij, h}(t)$ have a jump. Letting $t_{1}<t_{2}<\cdots<t_{K_h}$ be the time-points with a jump, the estimation of $\boldsymbol{\beta}_{0,h}(t)$ involves solving $\boldsymbol{U}_h[\boldsymbol{\beta}_{h}(t),t]=\boldsymbol{0}$ for $t=t_{1},t_{2},\cdots,t_{K_h}$. The proposed estimates can be easily computed using off-the-shelf software that implements the generalized estimating equations approach.

\subsection{Asymptotic Properties}
\label{ss:properties}
In this section, we show that the proposed estimator $\boldsymbol{\hat{\beta}}_{n,h}$ is uniformly consistent and asymptotically Gaussian. We also propose a wild bootstrap approach for the computation of simultaneous confidence bands and the implementation of nonparametric hypothesis testing for the true functional regression coefficient $\{\boldsymbol{\beta}_{0,h}(t):t \in [0, \tau]\}$. The proofs of the results that follow are provided in the Supplementary Material. In this work, we assume the following regularity conditions.
\begin{itemize}
\item[C1.]
The model \eqref{eq1} is correctly specified and the true functional regression coefficient $\{\boldsymbol{\beta}_{0,h}(t):t \in [0, \tau]\}$ is a vector-valued cadlag function with $\sup _{t \in[0, \tau]}\|\boldsymbol{\beta}_{0,h}(t)\|<\infty$. 
\item[C2.]
The derivatives $\dot{f}_h(u)=\partial f_h(u)/\partial u$ of the inverse link function $f_h=g_h^{-1}$ are Lipschitz continuous on compact sets.
\item[C3.]
The cluster size $M$ is bounded, in the sense that there exists a constant $m_{0} \in \mathbb{N}_{+}$ such that $P(M \leq m_{0})=1$.
\item[C4.]
$S_{ij, h}(\cdot)$, $R_{ij, h}(\cdot)$, $\boldsymbol{X}_{ij}(\cdot)$, $i=1, \ldots, n$, $j=1, \ldots, M_{i}$, are cadlag functions and have total variations on $[0, \tau]$ bounded by some constant $c<\infty$ almost surely. Also, the total variation of $\tilde{Y}_{ij, h}(t)$ on $[0, \tau]$ has bounded second moment.
\item[C5.]
The infimum over $[0,\tau]$ of the miminum eigenvalue $\inf_{t \in[0, \tau]} \text{eigmin } E [M^{-1} \sum_{j=1}^{M}I(M \leq m_0)S_{j, h}(t) R_{j, h}(t) \boldsymbol{X}_{j, h}(t) \boldsymbol{X}_{j, h}^{T}(t)]$ is strictly positive.
\item[C6.]
The class of functions $\{V_{j,h}(\boldsymbol{b}, t): \boldsymbol{b} \in B, t \in[0, \tau]\}$ is bounded above and below by positive constants, has bounded uniform entropy integral with bounded envelope, and is pointwise measurable for any bounded set $B \subset \mathbb{R}^{p}$. Also, the map $\boldsymbol{b}\mapsto V_{j,h}(\boldsymbol{b}, t)$, $t\in[0,\tau]$, is Lipschitz continuous.
\item[C7.]
$S_{ij, h}(\cdot)$, $R_{ij, h}(\cdot)$, $\boldsymbol{X}_{ij}(\cdot)$ and $\tilde{Y}_{ij, h}(\cdot)$ are identically distributed conditionally on the cluster size $M_i$, in the sense that $E[\tilde{Y}_{ij, h}(t)|M_i]=E[\tilde{Y}_{i1, h}(t)|M_i]$, $E[S_{ij, h}(t)|M_i]=E[S_{i1, h}(t)|M_i]$, $E[R_{ij, h}(t)|M_i]=E[R_{i1, h}(t)|M_i]$, and $E[\boldsymbol{X}_{ij}(t)|M_i]=E[\boldsymbol{X}_{i1}(t)|M_i]$, for all $i=1,\ldots,n$, $j=1,\ldots,m_0$, and $t\in[0,\tau]$.
\end{itemize}

Conditions C1, C2, and C4 - C6 were previously imposed in the temporal process regression framework by \citet{fine2004temporal}. 
The additional conditions C3 and C7 are imposed to account for the within-cluster dependence and the ICS. Both conditions are realistic in real-world multicenter studies and clinical trials. The following theorem states the uniform consistency of the proposed estimator $\{\boldsymbol{\hat{\beta}}_{n,h}(t):t \in [0, \tau]\}$. 

\begin{thm}\label{th1}
If regularity conditions C1 - C7 hold, then 
$$
\sup_{t\in [0,\tau]}\left\|\boldsymbol{\hat{\beta}}_{n,h}(t)-\boldsymbol{\beta}_{0,h}(t)\right\|\overset{as*}\rightarrow 0,
$$ 
as $n\rightarrow\infty$.
\end{thm}
The next theorem states that the proposed estimator is asymptotically Gaussian. This theorem provides the basis for conducting pointwise and simultaneous inference about the functional regression coefficients. Before providing this theorem, we define the influence function of the proposed estimator
$$
\boldsymbol{\psi}_{ij,h}(t)=\left[\boldsymbol{H}_{h}(t)\right]^{-1} \boldsymbol{A}_{ij,h}\left[\boldsymbol{\beta}_{0,h}(t),t\right],
$$ where
$$
\boldsymbol{H}_{h}(t)=E\left\{M^{-1} \sum_{j=1}^{M}S_{j, h}(t)R_{j, h}(t)\boldsymbol{D}_{j, h}^{T}\left[\boldsymbol{\beta}_{0,h}(t)\right]V_{j, h}\left[\boldsymbol{\beta}_{0,h}(t),t\right]\boldsymbol{D}_{j, h}\left[\boldsymbol{\beta}_{0,h}(t)\right]\right\}.
$$
Also, let $\psi_{ij,hl}(t)$ denote the ($l+1$)th component of (the vector-valued) $\boldsymbol{\psi}_{ij,h}(t)$, $l=0,1,\ldots,p$, which corresponds to the ($l+1$)th component of the covariate vector $\boldsymbol{X}(t)$. The empirical versions of $\boldsymbol{\psi}_{ij,h}(t)$ and $\psi_{ij,hl}(t)$ are denoted by $\hat{\boldsymbol{\psi}}_{ij,h}(t)$ and $\hat{\psi}_{ij,hl}(t)$, respectively. In addition, we let $\{l_{c}^{\infty}[0, \tau]\}^p$ denote the space of vector-valued real functions defined on $[0, \tau]$ with absolute value bounded above by $c$. Finally, let $\{\xi_{i}\}_{i=1}^{n}$ denote a random sample of standard normal variables which is independent of the observed data.

\begin{thm}\label{th2}
If regularity conditions C1 - C7 hold, then
$$
\sqrt{n}\left[\boldsymbol{\hat{\beta}}_{n,h}(t)-\boldsymbol{\beta}_{0,h}(t)\right]=\frac{1}{\sqrt{n}} \sum_{i=1}^{n} \frac{1}{M_i} \sum_{j=1}^{M_{i}} \boldsymbol{\psi}_{i j,h}(t)+o_{p}(1), \quad t\in[0,\tau],
$$ 
for $h \in \mathcal{T}^c$, and the class of influence functions $\{M^{-1}\sum_{j=1}^M\boldsymbol{\psi}_{j,h}(t):t\in[0,\tau]\}$ is Donsker. In addition, ${\hat{W}}_{n,hl}(\cdot)=n^{-1/2} \sum_{i=1}^{n}[ M_{i}^{-1} \sum_{j=1}^{M_{i}} \hat{\psi}_{i j,hl}(\cdot)] \xi_{i}$ converges weakly, conditionally on the observed data, to the same limiting process as the sequence $\sqrt{n}[\hat{\beta}_{n,hl}(\cdot)-\beta_{0,hl}(\cdot)]$, where $\hat{\beta}_{n,hl}(\cdot)$ and $\beta_{0,hl}(\cdot)$ are the ($l+1$)th components of $\boldsymbol{\hat{\beta}}_{n,h}(\cdot)$ and $\boldsymbol{\beta}_{0,h}(\cdot)$, respectively.
\end{thm}
By the latter theorem, $\sqrt{n}[\boldsymbol{\hat{\beta}}_{n,h}(\cdot)-\boldsymbol{\beta}_{0,h}(\cdot)]$ converges weakly to a tight mean zero Gaussian process $\boldsymbol{\mathbb{G}}_{h}(\cdot)$ in $\{l_{c}^{\infty}[0, \tau]\}^{p}$, with covariance function 
\[
\boldsymbol{\Sigma}_{h}(s, t)=E\left[\frac{1}{M}\sum_{j=1}^{M}\boldsymbol{\psi}_{j,h}(s)\right]\left[\frac{1}{M} \sum_{j=1}^{M}\boldsymbol{\psi}_{j,h}(t)\right]^T, \quad s, t \in[0, \tau].
\]
A consistent estimator of this covariance function is 
$$
\boldsymbol{\hat{\Sigma}}_{h}(s, t)=\frac{1}{n} \sum_{i=1}^{n}\left[\frac{1}{M_{i}} \sum_{j=1}^{M_i}\boldsymbol{\hat{\psi}}_{i j,h}(s)\right]\left[ \frac{1}{M_{i}} \sum_{j=1}^{M_i}\boldsymbol{\hat{\psi}}_{i j,h}(t)\right]^{T},\quad s,t \in [0, \tau].
$$ 
Explicit formulas for the empirical versions of the influence functions $\hat{\boldsymbol{\psi}}_{ij,h}(t)$, $t\in[0,\tau]$, are provided in the Supplementary Material. For the calculation of simultaneous confidence bands for $\beta_{0,hl}(\cdot)$ we consider the weight function 
\[
\hat{q}_{hl}(t)=\left\{\frac{1}{n} \sum_{i=1}^{n}\left[\frac{1}{M_{i}} \sum_{j=1}^{M_i}\hat{\psi}_{i j,hl}(t)\right]^2\right\}^{-1/2},
\]
which is equal to the inverse of the estimated standard error of $\hat{\beta}_{n,hl}(t)$. Then, by Theorem \ref{th2}, it follows that $\hat{q}_{hl}(\cdot)\sqrt{n}[\hat{\beta}_{n,hl}(\cdot)-\beta_{0,hl}(\cdot)]$ and $\hat{q}_{hl}(\cdot)\hat{W}_{n,hl}(\cdot)$, conditionally on the observed data, have the same asymptotic distribution. Therefore, a $1-\alpha$ confidence band for $\beta_{0,hl}(t)$, $t\in[t_1,t_2]\subset[0,\tau]$, can be computed as
$$
\hat{\beta}_{n,hl}(t) \pm \frac{c_{\alpha, hl}}{\sqrt{n} \hat{q}_{hl}(t)}, \quad t \in[t_{1}, t_{2}],
$$ 
where $c_{\alpha, hl}$ is the $1-\alpha$ empirical percentile of a sample of realizations of the random quantity $\sup _{t \in[t_{1}, t_{2}]}|\hat{q}_{hl}(t)\hat{W}_{n,hl}(t)|$. The latter sample of realizations can be obtained by repeatedly simulating sets of standard normal variables $\{\xi_i\}_{i=1}^n$ \citep{spiekerman1998marginal}. Algorithm \ref{alg:one} outlines the computation of the confidence bands. 

\begin{algorithm*}
\caption{Computation of a $1-\alpha$ confidence band for $\beta_{0,hl}(t)$, $t\in[t_1,t_2]$.}
\label{alg:one}
\BlankLine
\KwIn{observed data, $\alpha$}
\KwOut{$1-\alpha$ simultaneous confidence band}
\BlankLine

Calculate $\hat{q}_{hl}(t)$ and $\hat{\beta}_{n,hl}(t)$ \;
\tcp{Choose a large integer B (e.g., B = 1000)}
Choose $B \gets 1000$\; 
\BlankLine
\tcp{Simulate realizations from $\hat{q}_{hl}(t)\hat{W}_{n,hl}(t)$}
\For{$b=1 ,\dots, B$}{
Step 1 Simulate a set of independent standard normal random variables $\{\xi_i^{(b)}\}_{i=1}^{n}$\;
Step 2\ Calculate $\sup _{t \in[t_1, t_2]}|\hat{q}_{hl}(t)\hat{W}_{n,hl}^{(b)}(t)|$ based on $\{\xi_i^{(b)}\}_{i=1}^{n}$\;
}
\BlankLine
Calculate $c_{\alpha, hl} \gets 1-\alpha$ empirical percentile of the sample $\sup _{t \in[t_1, t_2]}|\hat{q}_{hl}(t)\hat{W}_{n,hl}^{(b)}(t)|$, $b=1,...,B$;
\BlankLine
\tcp{Calculate $1-\alpha$ confidence band}
Calculate $1-\alpha$ confidence band $\gets \hat{\beta}_{n,hl}(t) \pm {c_{\alpha, hl}}/[\sqrt{n} \hat{q}_{hl}(t)]$, $t \in[t_{1}, t_{2}]$\;
\BlankLine
\KwRet{$1-\alpha$ confidence band}
\end{algorithm*}

Since the confidence bands tend to be unstable at earlier and later time points, where there are fewer observed jumps of the response process, we suggest the restriction of the confidence band domain $[t_{1}, t_{2}]$ to the $5$th and $95$th or the $10$th and $90$th percentile of the observed time points where the response processes $\tilde{Y}_{ij,h}(\cdot)$ have a jump. 

Omnibus tests for nonparametric hypothesis testing regarding the covariate effects $\beta_{0,hl}(\cdot)$, $l=1,...,p$, can be conducted using a similar wild bootstrap approach. In many settings, the scientific question of interest is whether a given covariate is associated with the state occupation probability of the transient state of scientific interest. In such cases, the null hypothesis is $H_{0}: \beta_{0,hl}(\cdot)=0$, which indicates no covariate effect on the corresponding transient state occupation probability. The corresponding two-sided alternative hypothesis is $H_{1}: \beta_{0,hl}(\cdot)\neq 0$. In this setup, define the weighted Kolmogorov--Smirnov-type test statistic 
\[
K_{n,hl}=\sup _{t \in[0, \tau]}\left|\sqrt{n}\hat{q}_{hl}(t)\hat{\beta}_{n,hl}(t)\right|.
\]
The asymptotic distribution of $K_{n,hl}$ under the null hypothesis is quite complicated and, thus, we will utilize wild bootstrap to approximate this distribution and conduct hypothesis testing. By Theorem \ref{th2} and the continuous mapping theorem it follows that, simulation realizations from the asymptotic null distribution of $K_{n,hl}$ can be obtained by generating multiple sets of standard normal variables $\{\xi_{i}\}_{i=1}^n$ and, based on the latter sets, calculating multiple replications of the random quantity $\sup _{t \in[0, \tau]}|\hat{q}_{hl}(t)\hat{W}_{n,hl}(t)|$. The p-value can be estimated as the proportion of the latter replicates which are greater than or equal to the calculated value of the test statistic $K_{n,hl}$ based on observed data. Algorithm \ref{alg:two} summarizes the steps for computing p-values via the weighted Kolmogorov--Smirnov-type test.

\begin{algorithm*}
\caption{Nonparametric hypothesis testing procedure for $H_{0}: \beta_{0,hl}(t)=0$, $t \in[0, \tau]$.}
\label{alg:two}
\BlankLine
\KwIn{observed data}
\KwOut{$p$-value}
\BlankLine
\tcp{Calculate weighted Kolmogorov-Smirnov-type test statistic}
Calculate $K_{n,hl} \gets \sup _{t \in[0, \tau]}|\sqrt{n}\hat{q}_{hl}(t)\hat{\beta}_{n,hl}(t)|$ \;
\tcp{Choose a large integer B (e.g., B = 1000)}
Choose $B \gets 1000$\; 
\BlankLine
\tcp{Simulate realizations from the null hypothesis}
\For{$b=1 ,\dots, B$}{
Step 1 Simulate a set of independent standard normal random variables $\{\xi_i^{(b)}\}_{i=1}^{n}$\;
Step 2\ Calculate $\sup _{t \in[0, \tau]}|\hat{q}_{hl}(t)\hat{W}_{n,hl}^{(b)}(t)|$ based on $\{\xi_i^{(b)}\}_{i=1}^{n}$;
}
\BlankLine
\tcp{Calculate $p$-value}
Calculate $p$-value $\gets B^{-1}\sum_{b=1}^{B}I[\sup _{t \in[0, \tau]}|\hat{q}_{hl}(t)\hat{W}_{n,hl}^{(b)}(t)| \geq K_{n,hl}]$\;
\BlankLine
\KwRet{$p$-value}
\end{algorithm*}

To enhance the performance of the test with small numbers of clusters, we suggest the restriction of the comparison interval $[0,\tau]$ to the subinterval with limits the $5$th and $95$th or the $10$th and $90$th percentile of the observed time points where the response processes $\tilde{Y}_{ij,h}(\cdot)$ have a jump. This restriction is similar to the one suggested above for the confidence bands.

\section{Simulation Studies}
\label{s:sim}
A series of simulation studies were conducted to evaluate the finite sample performance of the proposed estimator and the nonparametric test for the covariate effects. We considered a study with clustered observations from a multistate process with three states $\mathcal{S}=\{1,2,3\}$, with absorbing state subspace $\mathcal{T}=\{3\}$, under ICS. We also simulated a covariate vector $\boldsymbol{X}=(1,X_{1}, X_{2})^T$, where $X_{1} \sim \textrm{Normal}(0,2^2)$ and $X_{2} \sim \textrm{Normal}(0,3^2)$. To induce within-cluster dependence, we generated cluster-specific random effects for each state $\mathcal{S}=\{1,2,3\}$. The times to the absorbing state (state 3) were generated under a Cox proportional hazards shared frailty model with positive stable frailty \citep{hougaard1986class, cong2007marginal, liu2011positive} of the form $\lambda(t|\gamma_i,\boldsymbol{X})=\gamma_i\lambda_0(t)\exp(\beta_{0,3,1}^{'}X_{1}+\beta_{0,3,2}^{'}X_{2})$, where $\gamma_i \sim \text{postive stable}(\alpha)$ with $\alpha=0.5$, $\lambda_0(t)=1$, and $(\beta_{0,3,1}^{'},\beta_{0,3,2}^{'})=(-0.5,-0.5)$. The corresponding marginal model is $\lambda(t|\boldsymbol{X})=\alpha\lambda_0(t)\Lambda_0(t)^{\alpha-1}\exp(\alpha\beta_{0,3,1}^{'}X_{1}+\alpha\beta_{0,3,2}^{'}X_{2})$, where $\Lambda_0(t)=\int_{0}^{t} \lambda_{0}(s)ds$. The response processes for states 1 and 2, conditional on being alive (i.e., not in state 3), were generated on fixed grid time points from a random intercept logistic regression model, with random effects following bridge distribution \citep{wang2003matching}, of the form $\operatorname{logit}\{E[\tilde{Y}_{2}(t)|b_i,\boldsymbol{X}, S_2(t)=1]\}=b_i+\beta_{0,2,0}^{'}(t)+\beta_{0,2,1}^{'}X_{1}+\beta_{0,2,2}^{'}X_{2}$, where $b_i \sim \text{Bridge}(0,\phi)$ with $\phi=0.5$, $\beta_{0,2,0}^{'}(t)=\log(t)$, and $(\beta_{0,2,1}^{'},\beta_{0,2,2}^{'})=(-0.5,-0.5)$. The corresponding marginal model is
\begin{equation}
\operatorname{logit}\{E[\tilde{Y}_{h}(t)|\boldsymbol{X}, S_{h}(t)=1]\}=\beta_{0,h0}(t)+\beta_{0,h1}X_{1}+\beta_{0,h2}X_{2}, \quad h=1,2,
\label{eq4}
\end{equation}
where $\beta_{0,h0}(t)=\phi\beta_{0,h0}^{'}(t)$, $\beta_{0,h1}=\phi\beta_{0,h1}^{'}$, and $\beta_{0,h2}=\phi\beta_{0,h2}^{'}$. In this simulation study, $S_1(t)=S_2(t)=1$ if an individual is not in the absorbing state (state 3) time $t$ and $S_1(t)=S_2(t)=0$ otherwise. The (random) right censoring times were independently generated from the $\textrm{Exp} (0.1)$ distribution and we also considered the maximum follow-up time $\tau=2$ (administrative right censoring time). These choices led to a 27.7\% right censoring rate on average. 

In the simulation studies, we considered scenarios with $n=50$, 100, 200, and 400 clusters. To induce ICS, the cluster sizes $M_i$, $i=1,\dots,n$, were generated from a mixture of discrete uniform distributions $\mathcal{U}(20,30)$, $\mathcal{U}(30,50)$ and $\mathcal{U}(50,60)$ with $M_i \sim \mathcal{U}(20,30)$ if $\gamma_{i}<\textrm{median}(\gamma)$ and $b_{i}<\textrm{median}(b)$,
$M_i \sim \mathcal{U}(50,60)$ if $\gamma_{i} \ge \textrm{median}(\gamma)$ and $b_{i}\ge \textrm{median}(b)$, and $M_i \sim \mathcal{U}(30,50)$, otherwise. For each simulation setting, we simulated 1000 datasets, and analyzed each dataset using the proposed method and the previously proposed method which does not account for the within-cluster dependence and the ICS \citep{fine2004temporal}. In this simulation study, we focused on the transient state 2, conditional on being alive, and all analyses were conducted using the marginal functional logistic model \eqref{eq4}, which was correctly specified in all settings. The weight function used in both analytical approaches was $V_{2}[\boldsymbol{\beta}_{2}(t),t]={1}/\{g_2^{-1}[\boldsymbol{\beta}_{2}^{T}(t)\boldsymbol{X}(t)]\{1-g_2^{-1}[\boldsymbol{\beta}_{2}^{T}(t)\boldsymbol{X}(t)]\}\},
$
where the inverse of link function is $g_{2}^{-1}(\theta)=\exp(\theta)/[1+\exp(\theta)]$. The standard errors were estimated using the corresponding closed-form estimators. The $95\%$ simultaneous confidence bands for $\beta_{0,2l}(t)$, $l=0,1,2$, and the p-values from the nonparametric Kolmogorov--Smirnov-type tests were computed based on 1000 simulated realizations of standard normal variables $\left\{\xi_{i}\right\}_{i=1}^{n}$, according to the Algorithms \ref{alg:one} and \ref{alg:two} (presented in Section~\ref{ss:properties}). The limits of the time domain $[t_1,t_2]$ for the confidence bands and the Kolmogorov--Smirnov-type tests were chosen to be the $10\%$ and $90\%$ percentile of the observed times where the response processes $\tilde{Y}_{ij,2}(\cdot)$ had a jump.

The simulation results for the pointwise estimates of the functional regression parameters $\beta_{0,2,1}$ are summarized in Table~\ref{t:one}. The corresponding results for the regression parameters $\beta_{0,2,0}$ and $\beta_{0,2,2}$ are provided in Tables 1 and 2 of the Supplementary Material. The proposed estimators were approximately unbiased, and the averages of the proposed standard error estimates were close to the Monte Carlo standard deviations of the estimates. This provides numerical evidence for the consistency of our proposed estimators for the regression parameters $\boldsymbol{\hat{\beta}}_{n,h}(t)$ as well as their associated standard error estimator. The empirical coverage probabilities were close to the nominal level in all settings. In contrast, the previous method for independent observations \citep{fine2004temporal}, provided standard error estimates that were smaller than the corresponding Monte Carlo standard deviations of the estimates. This under-estimation of the standard errors is attributed to the within-cluster dependence. As expected, ignoring the within-cluster dependence and the ICS resulted in poor coverage probabilities of the corresponding pointwise $95\%$ confidence intervals.

Results regarding the empirical coverage probabilities of the $95\%$ simultaneous confidence bands are presented in Table~\ref{t:two}. The proposed $95\%$ simultaneous confidence bands had coverage probabilities close to the nominal level for sufficiently large cluster sizes. In contrast, ignoring the within-cluster dependence and the ICS resulted in $95\%$ simultaneous confidence bands with a poor coverage. Finally, simulation results about the empirical rejection rates based on the proposed nonparametric Kolmogorov--Smirnov-type tests are presented in Table~\ref{t:three}. Under the null hypothesis $H_0: \beta_{0,2l}(\cdot)=0$, $l=1,2$, the empirical type I error rates of the proposed test were close to the nominal level $\alpha= 0.05$, with a sufficiently large number of clusters. In addition, the empirical power levels were increasing with the number of clusters $n$, which provides numerical evidence for the consistency of the proposed Kolmogorov--Smirnov-type test. 

Simulation results under a more variable cluster size $M\in\{5,\ldots,105\}$, are presented in Tables 3--7 of the Supplementary Material. This simulation setting induced a more pronounced ICS situation. The performance of the proposed methodology in this case remained satisfactory. In contrast, the previous method for independent observations \citep{fine2004temporal} provided biased estimates, more severely under-estimated standard errors, and poorer coverage probabilities of the 95\% pointwise confidence intervals and bands, compared to the previous simulation setting with less variable cluster size $M\in\{20,\ldots,60\}$. To sum up, our simulation studies showed that the proposed method performs well in finite samples, and that ignoring the within-cluster dependence and the ICS leads to invalid inferences. 

\section{SPECTRUM Trial Data Analysis}
\label{s:data}
\begin{figure}
\begin{center}
 \includegraphics[width=0.7\textwidth]{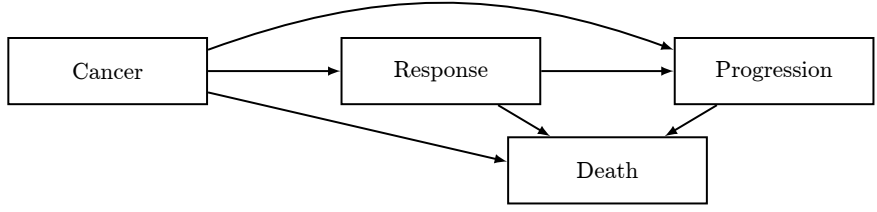}
\caption{An illustration of the multistate process in the SPECTRUM trial.}
\label{f:one}
\end{center}
\end{figure}

The proposed methodology was applied to analyze the data from the SPECTRUM trial, a multicenter phase III randomized trial on recurrent or metastatic squamous cell carcinoma of the head and neck \citep{vermorken2013cisplatin}. The treatments under comparison were chemotherapy combined with panitumumab and chemotherapy alone. The event history in this trial included the clinical states “initial cancer state”, “tumor response”, “disease progression", and “death” (Figure~\ref{f:one}). The randomization in this trial was stratified according to ECOG performance status (fully active vs other), prior treatment history (recurrent vs newly diagnosed), and site of primary tumor (oropharynx/larynx vs hypopharynx/oral cavity). Other important baseline covariates included sex, race, and age in years at screening. In total, 520 patients from $n=93$ clinics (i.e., clusters) were included in the dataset. Among these patients, 260 were included in the chemotherapy + panitumumab arm while 260 were included in the chemotherapy alone arm. The cluster sizes $M_i$, $i=1,\ldots,93$, ranged from 1 to 23 patients, with the median (interquartile range) being 4 (2, 8) patients. During the study, 138 patients (80 in the chemotherapy + panitumumab arm and 58 in the chemotherapy alone arm) achieved a tumor response. Moreover, 397 patients (197 in the chemotherapy + panitumumab arm and 200 from the chemotherapy alone arm) experienced disease progression. In addition, 425 patients (221 in the chemotherapy + panitumumab arm and 204 from the chemotherapy alone arm) died during the follow-up period. Among the 425 patients who died during the study, 56 patients (36 in the chemotherapy + panitumumab arm and 20 from the chemotherapy alone arm) died while being in the initial cancer state, 365 patients (182 in the chemotherapy + panitumumab arm and 183 from the chemotherapy alone arm) died while in the disease progression state, and 4 patients (3 in the chemotherapy + panitumumab arm and
1 from the chemotherapy alone arm) died while in the tumor response state. Finally, 95 patients (39 in the chemotherapy + panitumumab arm and 56 in the chemotherapy alone arm) were either lost to follow-up or were alive at the end of the study (right-censored individuals). Descriptive characteristics of the patients included in this analysis are presented in Table~\ref{t:four}.

In this analysis, we focused on the probability of being in the "tumor response" state, defined as a significant shrinkage of the tumor lesions per RECIST 1.0 criteria \citep{RECIST}, among those alive. We considered the following model for the estimation of the unadjusted treatment effect:
$$
\operatorname{logit}\left\{E[\tilde{Y}_{2}(t)|\boldsymbol{X}(t), S_{2}(t)=1]\right\}=\beta_{0,2,0}(t)+\beta_{0,2,1}(t)I(Trt=1),
$$
where $Trt$ represents the treatment arm with 1 denoting the chemotherapy + panitumumab arm and 0 denoting the chemotherapy alone arm. In this analysis, $S_1(t)=S_2(t)=1$ if the individual is alive at time $t$ and $S_1(t)=S_2(t)=0$, otherwise. We also estimated the treatment effect adjusting for the stratification variables (ECOG performance status, prior treatment history, and site of primary tumor) and
some other important baseline covariates (sex, race, and age), under the following model:
\begin{eqnarray}
\operatorname{logit}\left\{E[\tilde{Y}_{2}(t)|\boldsymbol{X}(t), S_{2}(t)=1]\right\}=\beta_{0,2,0}(t)+\beta_{0,2,1}(t)I(Trt=1) \nonumber\\
+ \beta_{0,2,2}(t)I(Sex=``male") + \beta_{0,2,3}(t)I(ECOG=1) + \beta_{0,2,4}(t)I(Ptrt = 1) \nonumber\\
+ \beta_{0,2,5}(t)I(TumorSite=1) + \beta_{0,2,6}(t)I(Race=1) + \beta_{0,2,7}(t)Age,\nonumber
\end{eqnarray}
where $ECOG$ is a binary variable for the ECOG performance status with 1 denoting the fully active status and 0 denoting the other statuses, $Ptrt$ represents the prior treatment history with 1 denoting patients with recurrent squamous-cell carcinoma of the head and neck and 0 denoting the newly diagnosed patients, $TumorSite$ represents the site of primary tumor with 1 denoting the oropharynx/larynx and 0 denoting the hypopharynx/oral cavity, and $Race$ is a binary variable with 1 indicating White or Caucasian race and 0 denoting the other races.

\begin{figure}[p]
\begin{center}
 \includegraphics[width=0.9\textwidth]{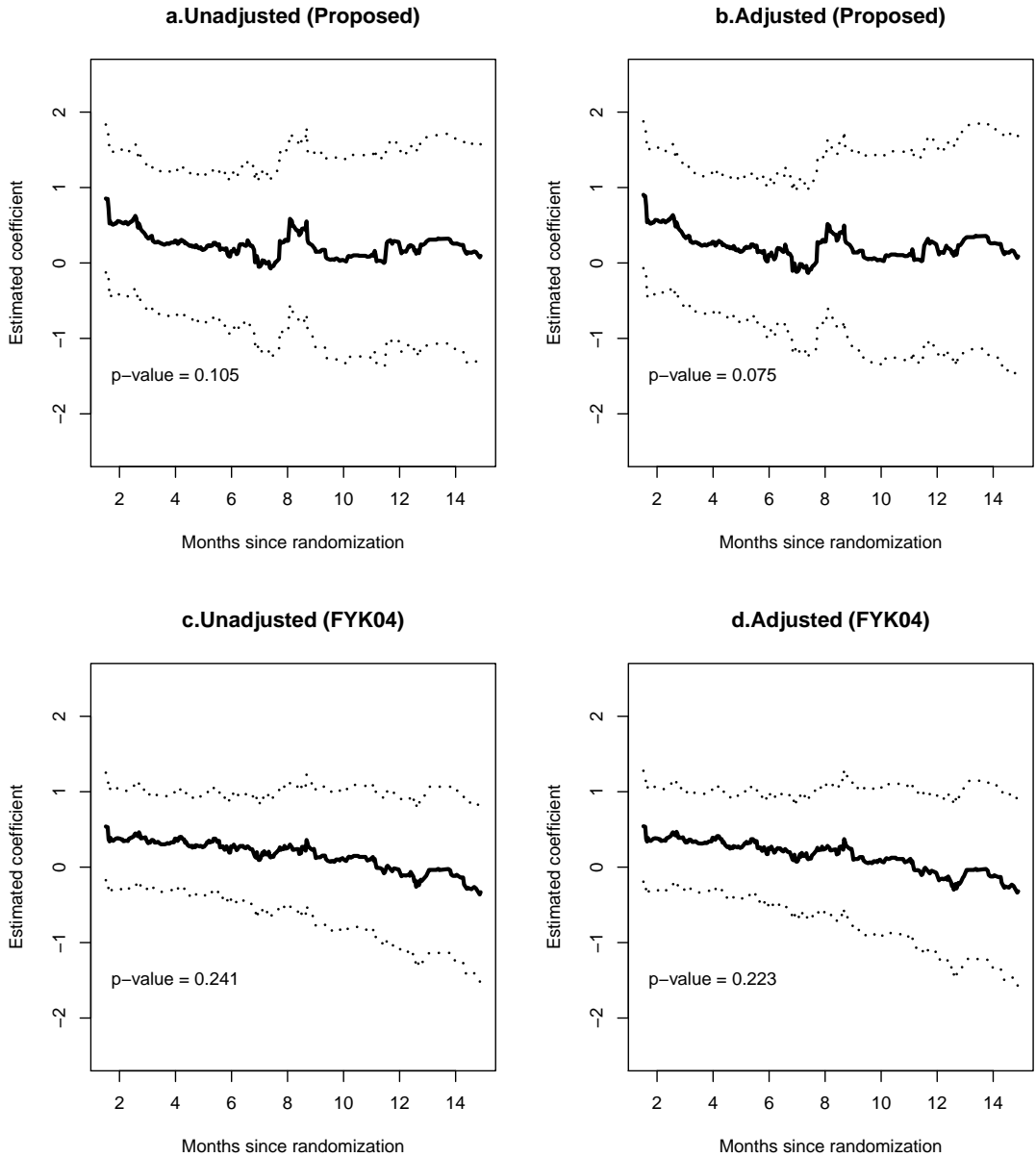}
\caption{Plots for the estimated regression coefficients of treatment effect along with the 95\% simultaneous confidence bands and the p-values based on the nonparametric tests for the tumor response state: (a) Unadjusted treatment effect based on the proposed method; (b) Adjusted treatment effect based on the proposed method; (c) Unadjusted treatment effect based on the method proposed by \citet{fine2004temporal} (FYK04) for independent observations; (d) Adjusted treatment effect based on the FYK04 for independent observations.}
\label{f:two}
\end{center}
\end{figure}

The estimated time-varying regression parameters for treatment (chemotherapy + panitumumab vs chemotherapy alone) along with the corresponding 95\% simultaneous confidence bands and the p-values (based on the proposed Kolmogorov--Smirnov-type tests) are presented in Figure~\ref{f:two}. For comparison, Figure~\ref{f:two} also includes the results from the analysis that ignores the within-cluster dependence and the ICS \citep{fine2004temporal}. Based on the proposed method, being in the panitumumab plus chemotherapy arm was associated with a higher probability of being in the "tumor response" state over time compared to the chemotherapy alone arm, conditionally on being alive. However, the unadjusted treatment effect was not statistically significant. After adjusting for the stratification variables and other potentially important baseline covariates (i.e., sex, ECOG performance status, prior treatment history, site of primary tumor, race, and age), the treatment effect became marginally significant (p-value=0.075). This indicates an efficiency increase as a result of adjusting for the stratification variables and other covariates. The analysis based on the previously proposed method \citep{fine2004temporal} provided somewhat less pronounced treatment effects, and this may be an indication of ICS. The 95\% simultaneous confidence bands from the latter approach were narrower compared to those from the proposed method, which is attributed to under-estimated standard errors as a consequence of ignoring the within-cluster dependence. 

The p-values for the remaining covariate effects in the adjusted model from the proposed method and the previously proposed method \citep{fine2004temporal} are summarized in Table~\ref{t:five}. Plots for the estimated time-varying regression coefficients, along with the corresponding 95\% simultaneous confidence bands and the p-values, are provided in the Supplementary Material. Based on the proposed method, the effects of sex and age on the state occupation probability of the tumor response state, conditionally on being alive, were statistically significant. In contrast, none of the covariate effects were statistically significant based on the previously proposed method for independent observations, which may be attributed to a potential ICS.

\section{Discussion}
\label{s:discuss}

In this paper, we addressed the issue of marginal regression analysis of transient state occupation probabilities with right-censored multistate process data with cluster-correlated observations, allowing also for ICS. To achieve this, we proposed a weighted functional generalized estimating equations approach. Rigorous methods for computing simultaneous confidence bands and a nonparametric hypothesis testing procedure for covariate effects were also proposed. Our methodology does not impose assumptions regarding the within-cluster dependence and does not require the Markov assumption. The validity of our methodology was justified both theoretically and via extensive simulation experiments. The latter experiments showed that ignoring the within-cluster dependence and the ICS can lead to invalid inferences. This highlights the practical importance of the proposed methodology in settings with clustered multistate processes and potential ICS. Software in the form of R code, together with a sample input data set and complete documentation, are available per request from the first author (wz11@iu.edu).

The issue of marginal analysis of clustered and right-censored multistate process data has received some attention in the recent literature. \citet{bakoyannis2021nonparametric} proposed a nonparametric moment-based estimator and a two-sample Kolmogorov-Smirnov-type test for state occupation and transition probabilities. \citet{bakoyannis2022nonparametric} proposed additional nonparametric two-sample tests for this problem, which may be more powerful than the Kolmogorov-Smirnov-type test for certain alternative hypotheses. Even though these methods allow for ICS, they do not incorporate covariates. To the best of our knowledge, there was no method for marginal regression on transient state occupation probabilities for clustered and right-censored multistate process data prior to this work. Nevertheless, the problem of regression analysis of clustered multistate data is crucial in many applications, such as in our motivating SPECTRUM trial \citep{vermorken2013cisplatin}. The analysis of the data from the latter trial illustrated that, ignoring the within-cluster dependence and the potential ICS may lead to substantially biased inferences in practice. 

In situations with ICS, two populations are typically of interest: the population of \textit{all cluster members} (ACM) and the population of \textit{typical cluster members} (TCM) \citep{seaman2014methods}. The ACM population includes all observations from all clusters, while the TCM population is comprised of a randomly selected observation from each cluster. As a consequence, the larger clusters dominate the ACM population, while all clusters are equally represented in the TCM population. The choice of the most appropriate target population in a given study depends on the scientific question under investigation. More details about the practical relevance of the two populations can be found in the articles by \citet{bakoyannis2021nonparametric} and \citet{bakoyannis2022nonparametric}. In this work, the proposed methodology provides inferences for the TCM population, in an effort to adjust for the over-representation of larger clusters. However, in some applications, the ACM population may be more scientifically relevant, such as in public health and health policy studies. In such cases, the proposed methodology can be easily modified to provide inferences for the ACM population by simply removing the weight $1/M_i$ from the weighted functional generalized estimating equations and the empirical versions of the influence functions. 

There is a number of possible extensions to this work. For example, extending the methodology to incorporate incompletely observed covariates is both methodologically interesting and scientifically relevant in many studies \citep{chen2012marginal}. 
Another useful but challenging extension would be to relax the i.i.d. assumption across clusters and allow, for example, some dependence between clusters in close spatial proximity (e.g., hospitals from the same region). 

\section*{Acknowledgements}

The authors would like to thank Project Data Sphere (\url{www.projectdatasphere.org}) for permission to use the SPECTRUM data. Neither Project Data Sphere nor the owner(s) of any information from the website have contributed to, approved, or are in any way responsible for the contents of this article.

\section*{Funding}

This research was supported in part by the National Institutes Health grant number R21AI145662 and Lilly Endowment, Inc., through its support for the Indiana University Pervasive Technology Institute.

\section*{Supplement}
\subsection*{Supplementary Material A}
Proofs of Theorem~\ref{th1}, Theorem~\ref{th2} and explicit formulas for the empirical versions of the influence functions referred in Section~\ref{s:method} are provided in Supplementary Material A.

\subsection*{Supplementary Material B}
Additional simulation results and data analysis results referred in Section~\ref{s:sim} and Section~\ref{s:data} are provided in Supplementary Material B.

\begin{table}[p]
\singlespacing
\small
\caption{Pointwise simulation results for the estimated functional regression parameters $\hat{\beta}_{n,2,1}(\cdot)$ based on the proposed approach and the approach by \citet{fine2004temporal} (FYK04) for independent observations.}
\label{t:one}
\begin{center}
\begin{tabular}{llcccccccc}
\hline
& & \multicolumn{4}{c}{Proposed} & \multicolumn{4}{c}{FYK04}\\
$n$ & $t$ & Bias & ASE & MCSD & CP & Bias & ASE & MCSD & CP\\ 
\hline
50	&0.2	&-0.003	&0.053	&0.057	&0.924	&-0.013	&0.040	&0.057	&0.818\\
	&0.4	&-0.003	&0.056	&0.058	&0.940	&-0.013	&0.043	&0.057	&0.855\\
	&0.6	&-0.003	&0.058	&0.062	&0.928	&-0.014	&0.045	&0.062	&0.845\\
	&0.8	&-0.005	&0.061	&0.064	&0.935	&-0.015	&0.048	&0.063	&0.862\\
	&1.0	&-0.004	&0.063	&0.069	&0.924	&-0.015	&0.050	&0.068	&0.840\\
	&1.2	&-0.004	&0.065	&0.071	&0.924	&-0.015	&0.053	&0.068	&0.861\\
	&1.4	&-0.007	&0.067	&0.073	&0.926	&-0.017	&0.055	&0.072	&0.862\\
	&1.6	&-0.004	&0.070	&0.076	&0.916	&-0.015	&0.057	&0.074	&0.870\\
	&1.8	&-0.004	&0.072	&0.078	&0.915	&-0.015	&0.059	&0.075	&0.872\\
100	&0.2	&-0.002	&0.038	&0.038	&0.944	&-0.013	&0.028	&0.038	&0.822\\
	&0.4	&-0.002	&0.040	&0.041	&0.950	&-0.013	&0.030	&0.040	&0.834\\
	&0.6	&-0.004	&0.042	&0.041	&0.946	&-0.015	&0.032	&0.041	&0.843\\
	&0.8	&-0.004	&0.044	&0.043	&0.956	&-0.015	&0.034	&0.043	&0.863\\
	&1.0	&-0.005	&0.045	&0.046	&0.949	&-0.015	&0.035	&0.045	&0.865\\
	&1.2	&-0.005	&0.047	&0.048	&0.941	&-0.016	&0.037	&0.047	&0.863\\
	&1.4	&-0.005	&0.048	&0.048	&0.958	&-0.015	&0.038	&0.047	&0.880\\
	&1.6	&-0.005	&0.050	&0.050	&0.938	&-0.015	&0.040	&0.049	&0.884\\
	&1.8	&-0.005	&0.051	&0.053	&0.944	&-0.016	&0.041	&0.052	&0.880\\
200	&0.2	&-0.002	&0.027	&0.027	&0.952	&-0.013	&0.020	&0.027	&0.818\\
	&0.4	&-0.002	&0.028	&0.028	&0.951	&-0.013	&0.021	&0.027	&0.831\\
	&0.6	&-0.002	&0.030	&0.030	&0.949	&-0.013	&0.022	&0.029	&0.824\\
	&0.8	&-0.001	&0.031	&0.032	&0.936	&-0.012	&0.024	&0.031	&0.840\\
	&1.0	&-0.001	&0.032	&0.032	&0.947	&-0.012	&0.025	&0.032	&0.845\\
	&1.2	&-0.002	&0.033	&0.035	&0.935	&-0.013	&0.026	&0.034	&0.835\\
	&1.4	&-0.002	&0.034	&0.034	&0.951	&-0.013	&0.027	&0.034	&0.869\\
	&1.6	&-0.002	&0.036	&0.036	&0.947	&-0.013	&0.028	&0.035	&0.860\\
	&1.8	&-0.002	&0.037	&0.037	&0.941	&-0.012	&0.029	&0.036	&0.880\\
400	&0.2	&-0.001	&0.019	&0.018	&0.969	&-0.011	&0.014	&0.018	&0.795\\
	&0.4	&-0.001	&0.020	&0.020	&0.943	&-0.011	&0.015	&0.020	&0.799\\
	&0.6	&-0.001	&0.021	&0.021	&0.950	&-0.012	&0.016	&0.021	&0.799\\
	&0.8	&-0.001	&0.022	&0.022	&0.942	&-0.012	&0.017	&0.022	&0.797\\
	&1.0	&-0.001	&0.023	&0.023	&0.941	&-0.013	&0.017	&0.023	&0.825\\
	&1.2	&-0.001	&0.024	&0.022	&0.968	&-0.012	&0.018	&0.022	&0.851\\
	&1.4	&-0.001	&0.024	&0.024	&0.958	&-0.013	&0.019	&0.024	&0.833\\
	&1.6	&-0.001	&0.025	&0.025	&0.958	&-0.012	&0.020	&0.024	&0.834\\
	&1.8	&-0.002	&0.026	&0.025	&0.949	&-0.013	&0.021	&0.025	&0.849\\
\hline
\end{tabular}
\end{center}
{\footnotesize Note: $n$: number of clusters with the cluster size $M\in\{20,\ldots,60\}$; ASE: average estimated standard error; MCSD: Monte Carlo standard deviation of the estimates; CP: coverage probability of $95\%$ pointwise confidence interval.}
\end{table}

\begin{table}[p]
\caption{Simulation results for the coverage probabilities of the $95\%$ simultaneous confidence bands for the proposed method and the method by \citet{fine2004temporal} (FYK04) for independent observations.}
\label{t:two}
\begin{center}
\begin{tabular}{lcccccc}
\hline
& \multicolumn{3}{c}{Proposed} & \multicolumn{3}{c}{FYK04} \\
 $n$ & $\beta_{0,2,0}(t)$ & $\beta_{0,2,1}(t)$ & $\beta_{0,2,2}(t)$ & $\beta_{0,2,0}(t)$ & $\beta_{0,2,1}(t)$ & $\beta_{0,2,2}(t)$ \\
\hline
\\[0.1ex]
50	&0.921	&0.899	&0.883	&0.221	&0.813	&0.722\\
100	&0.932	&0.929	&0.933	&0.143	&0.801	&0.704\\
200 &0.943  &0.941  &0.934  &0.067  &0.795  &0.672\\
400	&0.949	&0.948	&0.954	&0.006	&0.776	&0.640\\
\hline\\
\end{tabular}
\end{center}
{\footnotesize Note: $n$: number of clusters with the cluster size $M\in\{20,\ldots,60\}$.}
\end{table}

\begin{table}[p]
\caption{Simulation results on the empirical type I error rates $(H_0)$ and the power levels $(H_1)$ for the proposed Kolmogorov--Smirnov-test and the corresponding test proposed by \citet{fine2004temporal} (FYK04) for independent observations, at the $\alpha=0.05$ level.}
\label{t:three}
\begin{center}
\begin{tabular}{lcccc}
\hline
 & \multicolumn{2}{c}{Proposed} & \multicolumn{2}{c}{FYK04} \\[1ex]
$n$ & $\beta_{0,2,1}(t)$ & $\beta_{0,2,2}(t)$ & $\beta_{0,2,1}(t)$ & $\beta_{0,2,2}(t)$\\[1ex]
\hline
 \multicolumn{5}{c}{$H_{0}: \beta_{0,2,l}(t)=0,\quad l=1,2$} \\[1ex]
50	&0.085	&0.081	&0.110	&0.180\\
100	&0.084	&0.067	&0.133	&0.187\\
200 &0.047  &0.059  &0.137  &0.201\\
400	&0.047	&0.057	&0.125	&0.201\\
\hline
\multicolumn{5}{c}{$H_{1}: \beta_{0,2,l}(t)=-0.05,\quad l=1,2$}\\[1ex]
50	&0.297	&0.388	&0.415	&0.608\\
100	&0.444	&0.610	&0.641	&0.838\\
200 &0.714  &0.861  &0.887  &0.976\\
400	&0.930	&0.991	&0.991	&1.000\\
\hline
\multicolumn{5}{c}{$H_{1}: \beta_{0,2,l}(t)=-0.1,\quad l=1,2$}\\[1ex]
50	&0.699	&0.843	&0.857	&0.958\\
100	&0.931	&0.987	&0.983	&1.000\\
200 &0.997  &1.000  &1.000  &1.000\\
400	&1.000	&1.000	&1.000	&1.000\\
\hline
\multicolumn{5}{c}{$H_{1}: \beta_{0,2,l}(t)=-0.25,\quad l=1,2$}\\[1ex]
50	&0.999	&0.998	&1.000	&1.000\\
100	&1.000	&1.000	&1.000	&1.000\\
200 &1.000  &1.000  &1.000  &1.000\\
400	&1.000	&1.000	&1.000	&1.000\\
\hline\\
\end{tabular}
\end{center}
{\footnotesize Note: $n$: number of clusters with the cluster size $M\in\{20,\ldots,60\}$.}
\end{table}

\begin{table}[p]
\small
\caption{SPECTRUM trial data analysis: Descriptive characteristics of the analysis sample.}
\label{t:four}
\begin{center}
\begin{tabular}{llll}
\hline
Variable & Panitumumab $+$ Chemotherapy & Chemotherapy Alone & Overall \\
 & ($N$=260) & ($N$=260) & ($N$=520) \\
\hline
&$n$ (\%) & $n$ (\%) & $n$ (\%) \\
Sex&&& \\
Male & 227 (87.3) & 228 (87.7) & 455 (87.5) \\
Female & 33 (12.7) & 32 (12.3) & 65 (12.5) \\
\\[0.1ex]
ECOG performance status&&& \\
Fully active & 190 (73.1) & 180 (69.2) & 370 (71.2) \\
Other & 70 (26.9) & 80 (30.8) & 150 (28.8) \\
\\[0.1ex]
Prior treatment history&&&\\
Recurrent & 206 (79.2) & 201 (77.3) & 407 (78.3) \\
Newly diagnosed& 54 (20.8) & 59 (22.7) & 113 (21.7) \\
\\[0.1ex]
Site of primary tumor&&&\\
Oropharynx/ Larynx& 154 (59.2) & 149 (57.3)& 303 (58.3) \\
Hypopharynx/ Oral cavity& 106 (40.8)& 111 (42.7) & 217 (41.7) \\
\\[0.1ex]
Race&&&\\
White or Caucasian & 236 (90.8) & 230 (88.5) & 466 (89.6) \\
Other & 24 (9.2)& 30 (11.5) & 54 (10.4) \\
\\
& Median (IQR) & Median (IQR) & Median (IQR) \\
\\
Age$^1$ & 58.0 (53.0, 62.0) & 58.5 (53.0, 63.0) & 58.0 (53.0, 63.0) \\
\hline\\
\end{tabular}
\end{center}
{\footnotesize Note: $^1$: Age in years at screening.}
\end{table}

\begin{table}[p]
\small
\caption{SPECTRUM trial data analysis: p-values for the functional covariate effects for the tumor response state based on the proposed Kolmogorov--Smirnov-test and the corresponding test proposed by \citet{fine2004temporal} (FYK04) for independent observations.}
\label{t:five}
\begin{center}
\begin{tabular}{lcc}
\hline
Variable & Proposed & FYK04\\
\hline
\\[0.1ex]
Treatment (panitumumab + chemotherapy = 1, chemotherapy alone = 0) & 0.075 & 0.223 \\
Sex (male = 1, female = 0) & 0.011 & 0.180 \\
ECOG performance status (fully active = 1, other = 0) & 0.251 & 0.361 \\
Prior treatment history (recurrent = 1, newly diagnosed = 0) & 0.271 & 0.159 \\
Site of primary tumor (oropharynx/larynx = 1, hypopharynx/ oral cavity = 0) & 0.630 & 0.632 \\
Race (White or Caucasian = 1, other = 0) & 0.518 & 0.878 \\
Age$^1$ (per year) & 0.044 & 0.775 \\
\hline\\
\end{tabular}
\end{center}
{\footnotesize Note: $^1$: Age in years at screening.}
\end{table}

\bibliographystyle{chicago}
\bibliography{biblio}

\appendix


\section{Asymptotic Theory Proofs}
\label{s:proofs}
We justify the asymptotic properties of the proposed estimators using empirical process theory techniques \citep{kosorok2008introduction, van1996weak}. In this section, we provide the proofs for Theorems 2.1 and 2.2, and the explicit formulas for the empirical versions of the influence functions. For notational simplicity, we omit the subscript $h$, indicating the transient state of interest, from the proofs.

\subsection{Proof of Theorem 2.1}

We first define 
$$
\boldsymbol{C}_{i}(\boldsymbol{\gamma}, \boldsymbol{\beta},  {t}) =\frac{1}{M_i} \sum_{j=1}^{M_i} {S}_{ij}(t)  {R}_{ij}(t)  \boldsymbol{D}_{ij}^{T}\left[\boldsymbol{\gamma}(t)\right]{V}_{ij}\left[\boldsymbol{\gamma}(t),  {t}\right]\left\{ \tilde{Y}_{ij}(t)- f\left[\boldsymbol{\beta}^{T}(t){X}_{ij}(t)\right]\right\},
$$
with $f=g^{-1}$ and $\boldsymbol{\gamma}, \boldsymbol{\beta} \in\left\{l_{c}^{\infty}[0, \tau]\right\}^{p}$, where $\{l_{c}^{\infty}[0, \tau]\}^p$ denotes the space of bounded vector-valued real functions defined on $[0, \tau]$ with absolute value bounded above by $c$. We omit the subindex $i$ and use the notation $\boldsymbol{C}(\boldsymbol{\gamma}, \boldsymbol{\beta},  {t})$ to denote the generic version of the random quanity $\boldsymbol{C}_{i}(\boldsymbol{\gamma}, \boldsymbol{\beta},  {t})$ for an arbitrary cluster. We first argue that the class of functions
$$
\mathcal{L}=\left\{ \boldsymbol{C}(\boldsymbol{\gamma}, \boldsymbol{\beta},  {t}): \boldsymbol{\gamma}, \boldsymbol{\beta} \in\left\{l_{c}^{\infty}[0, \tau]\right\}^{p}, t \in[0, \tau]\right\},
$$
is Donsker. By condition C3 we have that
\begin{eqnarray}
\boldsymbol{C}(\boldsymbol{\gamma}, \boldsymbol{\beta},  {t}) &=&\sum_{j=1}^{m_0} \frac{1}{M} I(M\geq j){S}_{j}(t)  {R}_{j}(t)  \boldsymbol{D}_{j}^{T}\left[\boldsymbol{\gamma}(t)\right]{V}_{j}\left[\boldsymbol{\gamma}(t),  {t}\right]\left\{ \tilde{Y}_{j}(t)- f\left[\boldsymbol{\beta}^{T}(t){X}_{j}(t)\right]\right\} \nonumber \\
&\equiv& \sum_{j=1}^{m_0} \frac{1}{M} I(M\geq j)\tilde{\boldsymbol{C}}_j(\boldsymbol{\gamma}, \boldsymbol{\beta},  {t}). \label{C_tilde}
\end{eqnarray}
Under conditions C1, C2, and C4-C6, and using arguments similar to those used in the proof of Theorem A1 in \citet{fine2004temporal}, it follows that the classes of functions 
$$
\left\{ \tilde{\boldsymbol{C}}_j(\boldsymbol{\gamma}, \boldsymbol{\beta},  {t}): \boldsymbol{\gamma}, \boldsymbol{\beta} \in\{l_{c}^{\infty}[0, \tau]\}^{p}, t \in[0, \tau]\right\}, \ \ \ \ j=1,\ldots,m_0,
$$
are Donsker. This result along with the fact that $E[M^{-1}I(M\geq j)]^2\leq 1$, $j=1,\ldots m_0$, and \eqref{C_tilde}, lead to the conclusion that the class $\mathcal{L}$ is also Donsker, since (finite) sums of products of Donsker classes with random variables with bounded second moments are Donsker. In addition to this result, we need to show that the proposed estimating equation is unbiased, that is 
\[
E[\boldsymbol{C}(\tilde{\boldsymbol{\beta}}, \boldsymbol{\beta}_0,  t)]=\boldsymbol{0},
\]
for all $\tilde{\boldsymbol{\beta}}\in\{l_c^{\infty}[0,\tau]\}^p$ and $t\in[0,\tau]$. By conditions C1, C3, C7, and the missing at random (MAR) assumption imposed in Section 2.1, we have, for all $\tilde{\boldsymbol{\beta}}\in\{l_c^{\infty}[0,\tau]\}^p$ and $t\in[0,\tau]$, that
\begin{eqnarray*}
&   & E\left[\boldsymbol{C}(\boldsymbol{\tilde{\beta}}, \boldsymbol{\beta}_{0}, t)\right]\\ 
& = & E\left[\frac{1}{M} \sum_{j=1}^{m_0}I(M \ge j)  S_{j}(t) R_{j}(t) \boldsymbol{D}_{j}^{T}\left[\boldsymbol{\tilde{\beta}}(t)\right]  V_{j}\left[\boldsymbol{\tilde{\beta}}(t),  t\right]\left\{\tilde{Y}_{j}(t)- f\left[\boldsymbol{\beta}_{0}^{T}(t) \boldsymbol{X}_{j}(t)\right]\right\}\right]\\
& = & E\left\{\frac{1}{M} \sum_{j=1}^{m_0}I(M \ge j) E \left[S(t) R(t) \boldsymbol{D}^{T}\left[\boldsymbol{\tilde{\beta}}(t)\right]  V\left[\boldsymbol{\tilde{\beta}}(t),  t\right]\left\{\tilde{Y}(t)- f\left[\left. \boldsymbol{\beta}_{0}^{T}(t) \boldsymbol{X}(t)\right]\right\}\right| M\right]\right\}\\
& = & E\left\{E \left[\left. S(t) R(t) \boldsymbol{D}^{T}\left[\boldsymbol{\tilde{\beta}}(t)\right]  V\left[\boldsymbol{\tilde{\beta}}(t),  t\right]\left\{\tilde{Y}(t)- f\left[\boldsymbol{\beta}_{0}^{T}(t) \boldsymbol{X}(t)\right]\right\}\right| M\right]\frac{1}{M} \sum_{j=1}^{m_0}I(M \ge j) \right\}\\
& = & E\left\{E \left[\left. S(t) R(t) \boldsymbol{D}^{T}\left[\boldsymbol{\tilde{\beta}}(t)\right]  V\left[\boldsymbol{\tilde{\beta}}(t),  t\right]\left\{\tilde{Y}(t)- f\left[\boldsymbol{\beta}_{0}^{T}(t) \boldsymbol{X}(t)\right]\right\}\right| M\right]\right\}\\
& = & E \left[S(t) R(t) \boldsymbol{D}^{T}\left[\boldsymbol{\tilde{\beta}}(t)\right]  V\left[\boldsymbol{\tilde{\beta}}(t),  t\right]\left\{\tilde{Y}(t)- f\left[\boldsymbol{\beta}_{0}^{T}(t) \boldsymbol{X}(t)\right]\right\}\right]\\
& = & E\left\{E\left[\left. S(t) R(t) \boldsymbol{D}^{T}\left[\boldsymbol{\tilde{\beta}}(t)\right] {V}\left[\boldsymbol{\tilde{\beta}}(t), t\right]\left\{\tilde{Y}(t)- f\left[\boldsymbol{\beta}_{0}^{T}(t) \boldsymbol{X}(t)\right]\right\} \right| \boldsymbol{X}(t),  S(t) = 1\right]\right\}\\
& = & E\left[\boldsymbol{D}^{T}\left[\boldsymbol{\tilde{\beta}}(t)\right] V\left[\boldsymbol{\tilde{\beta}}(t),  {t}\right] E\left[ R(t) | \boldsymbol{X}(t), S(t)=1\right] \left\{E\left[\left.\tilde{Y}(t) \right| \boldsymbol{X}(t), S(t)=1\right]- f\left[\boldsymbol{\beta}_{0}^{T}(t)  \boldsymbol{X}(t)\right]\right\}\right]\\
& = & \boldsymbol{0}.
\end{eqnarray*}
These facts and arguments similar to those used in the proof of Theorem A1 in \citet{fine2004temporal} conclude the proof of Theorem 2.1. 

\subsection{Proof of Theorem 2.2}

By definition, the proposed estimator satisfies
\begin{eqnarray*}
\boldsymbol{0} & = & n^{-1/2} \boldsymbol{U}[\boldsymbol{\hat{\beta}}_{n}(t), t]\\
& = & n^{-1/2} \boldsymbol{U}\left[\boldsymbol{\beta}_{0}(t), t\right] + n^{-1/2} \left\{\boldsymbol{U}[\boldsymbol{\hat{\beta}}_{n}(t), t] - \boldsymbol{U}\left[\boldsymbol{\beta}_{0}(t), t\right]\right\} \\
& = & n^{-1/2} \boldsymbol{U}\left[\boldsymbol{\beta}_{0}(t), t\right] + n^{-1/2} \sum_{i=1}^{n}\left[ \boldsymbol{C}_{i}(\boldsymbol{\hat{\beta}}_{n}, \boldsymbol{\beta}_{0},  t)- \boldsymbol{C}_{i}(\boldsymbol{\beta}_{0}, \boldsymbol{\beta}_{0},  t)\right] \\
&   & +n^{-1/2} \sum_{i=1}^{n} \frac{1}{M_i} \sum_{j=1}^{M_i} S_{ij}(t)  R_{ij}(t) \boldsymbol{D}_{ij}^{T}[\boldsymbol{\hat{\beta}}_{n}(t)] V_{ij}[\boldsymbol{\hat{\beta}}_{n}(t),  t]\left\{ f[\boldsymbol{\hat{\beta}}_{n}^{T}(t) \boldsymbol{X}_{ij}(t)]- f[\boldsymbol{\beta}_{0}^{T}(t) \boldsymbol{X}_{ij}(t)]\right\} \\
& = & \boldsymbol{U}_{0}(t)+\boldsymbol{U}_{1}(t)+\boldsymbol{U}_{2}(t)
\end{eqnarray*}

We will first argue that the $\sup_{t \in [0, \tau]}\|\boldsymbol{U}_{1}(t)\| \overset{p}\rightarrow 0$. First, note that the class of functions
\[
\left\{\boldsymbol{C}(\boldsymbol{\beta}, \boldsymbol{\beta}_0,  {t}) - \boldsymbol{C}(\boldsymbol{\beta}_0, \boldsymbol{\beta}_0,  {t}): \boldsymbol{\beta} \in\left\{l_{c}^{\infty}[0, \tau]\right\}^{p}, \sup_{t \in [0, \tau]}\|\boldsymbol{\beta}(t)-\boldsymbol{\beta}_0(t)\|<\delta,  t \in[0, \tau]\right\}
\]
is Donsker for some $\delta>0$, because it is formed by the difference between two classes which are subsets of the Donsker class $\mathcal{L}$. Next, let $C^{(l)}(\boldsymbol{\beta}, \boldsymbol{\beta}_{0},  t)$ denote the $(l+1)th$ component of $\boldsymbol{C}(\boldsymbol{\beta}, \boldsymbol{\beta}_0,  {t})$, $l=0,\ldots,p$, and $D_j^{(l)}(\boldsymbol{\beta}(t))$ be the $(l+1)th$ component of $\boldsymbol{D}_j(\boldsymbol{\beta}(t))$, $j=1,\ldots,m_0$. Under conditions C2-C4 and C6, and for any $t\in[0,\tau]$, we have that
\begin{eqnarray*}
E\left[ C^{(l)}(\boldsymbol{\beta}, \boldsymbol{\beta}_{0},  t) - C^{(l)}(\boldsymbol{\beta}_0, \boldsymbol{\beta}_{0},  t)\right]^2&\leq&E\left[\sum_{j=1}^{m_0}\left|D_j^{(l)}(\boldsymbol{\beta}(t))V_j(\boldsymbol{\beta}(t),t)-D_j^{(l)}(\boldsymbol{\beta}_0(t))V_j(\boldsymbol{\beta}_0(t),t)\right|\right]^2 \\
&\leq& m_0^2E\left[\max_{1\leq j \leq m_0}\left|D_j^{(l)}(\boldsymbol{\beta}(t))V_j(\boldsymbol{\beta}(t),t)-D_j^{(l)}(\boldsymbol{\beta}_0(t))V_j(\boldsymbol{\beta}_0(t),t)\right|\right]^2 \\
&\leq& \|\boldsymbol{\beta}(t)-\boldsymbol{\beta}_0(t)\|^2m_0^2K^2, \ \ \ \ l=0,\ldots,p,
\end{eqnarray*}
where $K<\infty$ is the supremum of the set of Lipschitz constants of the functions $\boldsymbol{\beta}(t)\mapsto D_j^{(l)}(\boldsymbol{\beta}(t))V_j(\boldsymbol{\beta}(t),t)$ for any covariate pattern in the covariate space and $t\in[0,\tau]$. This implies that  
\[
\sup _{t \in [0, \tau]} E\left[ C^{(l)}(\boldsymbol{\beta}, \boldsymbol{\beta}_{0},  t)-C^{(l)}(\boldsymbol{\beta}_0, \boldsymbol{\beta}_{0},  t)\right]^2\rightarrow 0, \ \ \ \ l=0,\ldots,p,
\]
as $\sup _{t \in [0, \tau]}\|\boldsymbol{\beta}(t)-\boldsymbol{\beta}_0(t)\|\rightarrow 0$. The latter two results along with Theorem 2.1 and arguments similar to those used in the proof of Lemma 3.3.5. in \citet{van1996weak} lead to the conclusion that 
\[
\sup_{t \in [0, \tau]}\left\|\boldsymbol{U}_{1}(t)\right\| \overset{p}\rightarrow 0.
\]
This result and arguments similar to those used in the proofs of theorems A2 and A3 in \citet{fine2004temporal} conclude the proof of Theorem 2.2. 

\section{Additional numerical Results}
\subsection{Additional Simulation Results}
Simulation results for the pointwise estimates of the regression parameters $\beta_{0,2,2}$ and $\beta_{0,2,0}$ are provided in Table~\ref{t:suppone} and Table~\ref{t:supptwo}, respectively.

\begin{table}[p]
\caption{Pointwise simulation results for the estimated functional regression parameters $\hat{\beta}_{n,2,2}(\cdot)$ based on the proposed approach and the approach by \citet{fine2004temporal} (FYK04) for independent observations.}
\label{t:suppone}
\begin{center}
\begin{tabular}{llcccccccc}
\hline
& & \multicolumn{4}{c}{Proposed} & \multicolumn{4}{c}{FYK04}\\
$n$ & $t$ & Bias & ASE & MCSD & CP & Bias  & ASE & MCSD & CP\\ 
\hline
50	&0.2	&-0.005	&0.046	&0.047	&0.924	&-0.016	&0.029	&0.048	&0.762\\
	&0.4	&-0.005	&0.048	&0.052	&0.922	&-0.015	&0.032	&0.052	&0.766\\
	&0.6	&-0.006	&0.05	&0.052	&0.918	&-0.017	&0.034	&0.053	&0.795\\
	&0.8	&-0.005	&0.052	&0.055	&0.925	&-0.015	&0.036	&0.055	&0.782\\
	&1.0	&-0.006	&0.054	&0.057	&0.922	&-0.016	&0.038	&0.057	&0.791\\
	&1.2	&-0.006	&0.056	&0.061	&0.916	&-0.017	&0.040	&0.060	&0.807\\
	&1.4	&-0.007	&0.058	&0.063	&0.933	&-0.017	&0.042	&0.062	&0.805\\
	&1.6	&-0.006	&0.059	&0.066	&0.911	&-0.016	&0.043	&0.065	&0.830\\
	&1.8	&-0.005	&0.061	&0.064	&0.923	&-0.015	&0.045	&0.064	&0.835\\
100	&0.2	&-0.003	&0.033	&0.033	&0.951	&-0.013	&0.021	&0.033	&0.728\\
	&0.4	&-0.003	&0.034	&0.036	&0.946	&-0.013	&0.022	&0.035	&0.763\\
	&0.6	&-0.003	&0.035	&0.036	&0.946	&-0.013	&0.024	&0.036	&0.785\\
	&0.8	&-0.003	&0.037	&0.039	&0.945	&-0.014	&0.025	&0.039	&0.768\\
	&1.0	&-0.004	&0.039	&0.040	&0.944	&-0.014	&0.027	&0.040	&0.793\\
	&1.2	&-0.004	&0.040	&0.041	&0.939	&-0.014	&0.028	&0.040	&0.792\\
	&1.4	&-0.004	&0.041	&0.042	&0.944	&-0.014	&0.029	&0.042	&0.818\\
	&1.6	&-0.004	&0.042	&0.044	&0.945	&-0.014	&0.030	&0.043	&0.823\\
	&1.8	&-0.004	&0.043	&0.045	&0.932	&-0.014	&0.032	&0.045	&0.822\\
200	&0.2	&-0.002	&0.023	&0.023	&0.945	&-0.013	&0.014	&0.024	&0.719\\
	&0.4	&-0.003	&0.024	&0.024	&0.947	&-0.013	&0.016	&0.024	&0.744\\
	&0.6	&-0.002	&0.025	&0.026	&0.950	&-0.013	&0.017	&0.025	&0.758\\
	&0.8	&-0.003	&0.026	&0.027	&0.940	&-0.014	&0.018	&0.027	&0.774\\
	&1.0	&-0.002	&0.027	&0.028	&0.949	&-0.013	&0.019	&0.027	&0.782\\
	&1.2	&-0.002	&0.028	&0.029	&0.939	&-0.013	&0.020	&0.029	&0.774\\
	&1.4	&-0.003	&0.029	&0.030	&0.944	&-0.014	&0.021	&0.030	&0.782\\
	&1.6	&-0.004	&0.030	&0.031	&0.942	&-0.014	&0.021	&0.031	&0.806\\
	&1.8	&-0.004	&0.031	&0.032	&0.942	&-0.014	&0.022	&0.032	&0.788\\
400	&0.2	&-0.001	&0.016	&0.016	&0.961	&-0.012	&0.010	&0.016	&0.685\\
	&0.4	&-0.001	&0.017	&0.017	&0.960	&-0.012	&0.011	&0.017	&0.689\\
	&0.6	&-0.002	&0.018	&0.018	&0.958	&-0.013	&0.012	&0.018	&0.718\\
	&0.8	&-0.001	&0.019	&0.019	&0.948	&-0.013	&0.013	&0.018	&0.706\\
	&1.0	&-0.001	&0.019	&0.019	&0.961	&-0.013	&0.013	&0.019	&0.735\\
	&1.2	&-0.002	&0.020	&0.020	&0.941	&-0.013	&0.014	&0.020	&0.742\\
	&1.4	&-0.002	&0.021	&0.020	&0.955	&-0.013	&0.014	&0.020	&0.760\\
	&1.6	&-0.002	&0.021	&0.022	&0.942	&-0.013	&0.015	&0.021	&0.758\\
	&1.8	&-0.002	&0.022	&0.022	&0.949	&-0.013	&0.016	&0.021	&0.778\\
\hline
\end{tabular}
\end{center}
{\footnotesize Note: $n$: number of clusters with cluster size $M\in\{20,\ldots,60\}$; ASE: average estimated standard error; MCSD: Monte Carlo standard deviation of the estimates; CP: coverage probability of $95\%$ pointwise confidence interval.}
\end{table}

\begin{table}[p]
\caption{Pointwise simulation results for the estimated functional regression parameters $\hat{\beta}_{n,2,0}(\cdot)$ based on the proposed approach and the approach by \citet{fine2004temporal} (FYK04) for independent observations.}
\label{t:supptwo}
\begin{center}
\begin{tabular}{llcccccccc}
\hline
& & \multicolumn{4}{c}{Proposed} & \multicolumn{4}{c}{FYK04}\\
$n$ & $t$ & Bias & ASE & MCSD & CP & Bias &  ASE & MCSD & CP\\ 
\hline
50	&0.2	&-0.013	&0.254	&0.253	&0.952	&0.225	&0.074	&0.257	&0.284\\
	&0.4	&-0.012	&0.258	&0.261	&0.950	&0.241	&0.084	&0.262	&0.294\\
	&0.6	&0.000	&0.267	&0.270	&0.944	&0.261	&0.095	&0.270	&0.314\\
	&0.8	&0.000	&0.276	&0.282	&0.946	&0.266	&0.105	&0.280	&0.357\\
	&1.0	&0.001	&0.287	&0.297	&0.938	&0.270	&0.115	&0.293	&0.375\\
	&1.2	&0.003	&0.296	&0.311	&0.930	&0.278	&0.125	&0.306	&0.379\\
	&1.4	&0.009	&0.307	&0.328	&0.926	&0.286	&0.134	&0.323	&0.402\\
	&1.6	&0.009	&0.315	&0.337	&0.929	&0.289	&0.143	&0.330	&0.426\\
	&1.8	&0.002	&0.325	&0.345	&0.929	&0.283	&0.152	&0.338	&0.458\\
100	&0.2	&0.000	&0.179	&0.183	&0.948	&0.240	&0.051	&0.183	&0.185\\
	&0.4	&0.002	&0.182	&0.192	&0.941	&0.257	&0.059	&0.190	&0.201\\
	&0.6	&0.006	&0.189	&0.199	&0.938	&0.270	&0.067	&0.196	&0.217\\
	&0.8	&0.009	&0.196	&0.205	&0.934	&0.279	&0.074	&0.202	&0.222\\
	&1.0	&0.013	&0.204	&0.217	&0.928	&0.287	&0.081	&0.212	&0.247\\
	&1.2	&0.011	&0.210	&0.224	&0.933	&0.289	&0.087	&0.218	&0.259\\
	&1.4	&0.015	&0.218	&0.233	&0.935	&0.298	&0.094	&0.228	&0.286\\
	&1.6	&0.014	&0.224	&0.240	&0.934	&0.298	&0.100	&0.235	&0.313\\
	&1.8	&0.013	&0.231	&0.246	&0.929	&0.299	&0.106	&0.242	&0.326\\
200	&0.2	&-0.005	&0.127	&0.130	&0.944	&0.236	&0.036	&0.133	&0.096\\
	&0.4	&-0.001	&0.129	&0.131	&0.943	&0.255	&0.042	&0.132	&0.095\\
	&0.6	&-0.003	&0.134	&0.136	&0.944	&0.262	&0.047	&0.136	&0.097\\
	&0.8	&0.002	&0.139	&0.144	&0.938	&0.273	&0.052	&0.143	&0.113\\
	&1.0	&0.000	&0.145	&0.147	&0.952	&0.277	&0.057	&0.146	&0.123\\
	&1.2	&0.002	&0.150	&0.154	&0.943	&0.282	&0.061	&0.151	&0.146\\
	&1.4	&0.004	&0.155	&0.155	&0.959	&0.287	&0.066	&0.153	&0.149\\
	&1.6	&0.005	&0.160	&0.166	&0.940	&0.291	&0.070	&0.163	&0.163\\
	&1.8	&0.007	&0.164	&0.171	&0.945	&0.294	&0.075	&0.169	&0.186\\
400	&0.2	&0.000	&0.090	&0.089	&0.957	&0.240	&0.026	&0.091	&0.016\\
	&0.4	&0.002	&0.091	&0.090	&0.950	&0.258	&0.029	&0.091	&0.018\\
	&0.6	&0.004	&0.095	&0.093	&0.956	&0.268	&0.033	&0.093	&0.014\\
	&0.8	&0.004	&0.099	&0.098	&0.947	&0.275	&0.037	&0.097	&0.018\\
	&1.0	&0.006	&0.102	&0.102	&0.945	&0.282	&0.040	&0.100	&0.016\\
	&1.2	&0.007	&0.106	&0.104	&0.944	&0.286	&0.043	&0.103	&0.029\\
	&1.4	&0.007	&0.110	&0.108	&0.949	&0.291	&0.047	&0.107	&0.027\\
	&1.6	&0.006	&0.113	&0.113	&0.950	&0.292	&0.050	&0.112	&0.032\\
	&1.8	&0.008	&0.117	&0.114	&0.955	&0.296	&0.053	&0.113	&0.038\\
\hline
\end{tabular}
\end{center}
{\footnotesize Note: $n$: number of clusters with cluster size $M\in\{20,\ldots,60\}$; ASE: average estimated standard error; MCSD: Monte Carlo standard deviation of the estimates; CP: coverage probability of $95\%$ pointwise confidence interval.}
\end{table}

Additional simulation results under a more variable cluster size $M\in\{5,\ldots,105\}$ are presented in Tables~\ref{t:suppthree}-\ref{t:suppseven}.

\begin{table}[p]
\caption{Pointwise simulation results for the estimated functional regression parameters $\hat{\beta}_{n,2,0}(\cdot)$ based on the proposed approach and the approach by \citet{fine2004temporal} (FYK04) for independent observations. Simulation scenario with a more variable cluster size $M\in\{5,\ldots,105\}$.}
\label{t:suppthree}
\begin{center}
\begin{tabular}{llcccccccc}
\hline
& & \multicolumn{4}{c}{Proposed} & \multicolumn{4}{c}{FYK04}\\
$n$ & $t$ & Bias & ASE & MCSD & CP & Bias  & ASE & MCSD & CP\\ 
\hline
50	&0.2	&-0.011	&0.255	&0.256	&0.946	&0.456	&0.065	&0.276	&0.101\\
	&0.4	&-0.003	&0.260	&0.264	&0.947	&0.513	&0.078	&0.274	&0.086\\
	&0.6	& 0.000	&0.270	&0.280	&0.948	&0.545	&0.089	&0.284	&0.083\\
	&0.8	& 0.011	&0.282	&0.292	&0.929	&0.578	&0.100	&0.293	&0.080\\
	&1.0	& 0.015	&0.293	&0.305	&0.943	&0.600	&0.111	&0.302	&0.099\\
	&1.2	& 0.016	&0.305	&0.327	&0.936	&0.620	&0.121	&0.315	&0.098\\
	&1.4	& 0.017	&0.314	&0.341	&0.927	&0.631	&0.131	&0.326	&0.119\\
	&1.6	& 0.013	&0.326	&0.343	&0.937	&0.644	&0.140	&0.336	&0.127\\
	&1.8	& 0.014	&0.337	&0.358	&0.935	&0.654	&0.150	&0.345	&0.138\\
100	&0.2	&-0.004	&0.180	&0.186	&0.947	&0.468	&0.046	&0.195	&0.025\\
	&0.4	& 0.001	&0.184	&0.194	&0.946	&0.523	&0.054	&0.197	&0.012\\
	&0.6	& 0.008	&0.192	&0.204	&0.937	&0.560	&0.062	&0.203	&0.012\\
	&0.8	& 0.011	&0.200	&0.212	&0.934	&0.587	&0.070	&0.209	&0.015\\
	&1.0	& 0.014	&0.209	&0.220	&0.933	&0.611	&0.077	&0.216	&0.017\\
	&1.2	& 0.018	&0.217	&0.231	&0.927	&0.627	&0.085	&0.225	&0.018\\
	&1.4	& 0.017	&0.224	&0.240	&0.928	&0.641	&0.091	&0.229	&0.016\\
	&1.6	& 0.014	&0.232	&0.247	&0.927	&0.651	&0.098	&0.237	&0.018\\
	&1.8	& 0.021	&0.240	&0.256	&0.931	&0.668	&0.104	&0.243	&0.022\\
200	&0.2	&-0.004	&0.127	&0.129	&0.954	&0.469	&0.032	&0.138	&0.004\\
	&0.4	&-0.002	&0.130	&0.132	&0.955	&0.521	&0.038	&0.138	&0.000\\
	&0.6	&-0.003	&0.136	&0.137	&0.947	&0.555	&0.044	&0.140	&0.000\\
	&0.8	&-0.001	&0.142	&0.144	&0.948	&0.581	&0.049	&0.145	&0.000\\
	&1.0	& 0.003	&0.148	&0.148	&0.950	&0.602	&0.055	&0.150	&0.000\\
	&1.2	& 0.002	&0.154	&0.157	&0.936	&0.618	&0.059	&0.156	&0.000\\
	&1.4	& 0.001	&0.160	&0.165	&0.946	&0.634	&0.064	&0.163	&0.000\\
	&1.6	&-0.002	&0.165	&0.169	&0.939	&0.641	&0.069	&0.166	&0.000\\
	&1.8	& 0.001	&0.171	&0.176	&0.944	&0.656	&0.073	&0.168	&0.000\\
400	&0.2	& 0.002	&0.090	&0.088	&0.951	&0.471	&0.023	&0.095	&0.000\\
	&0.4	& 0.002	&0.092	&0.091	&0.957	&0.522	&0.027	&0.096	&0.000\\
	&0.6	& 0.003	&0.096	&0.095	&0.958	&0.556	&0.031	&0.097	&0.000\\
	&0.8	& 0.006	&0.101	&0.098	&0.955	&0.584	&0.035	&0.100	&0.000\\
	&1.0	& 0.008	&0.105	&0.107	&0.946	&0.606	&0.039	&0.104	&0.000\\
	&1.2	& 0.006	&0.109	&0.109	&0.949	&0.620	&0.042	&0.108	&0.000\\
	&1.4	& 0.006	&0.113	&0.112	&0.949	&0.636	&0.045	&0.110	&0.000\\
	&1.6	& 0.006	&0.117	&0.114	&0.966	&0.647	&0.049	&0.110	&0.000\\
	&1.8	& 0.007	&0.121	&0.119	&0.948	&0.659	&0.052	&0.113	&0.000\\
\hline
\end{tabular}
\end{center}
{\footnotesize Note: $n$: number of clusters with cluster size $M\in\{5,\ldots,105\}$; ASE: average estimated standard error; MCSD: Monte Carlo standard deviation of the estimates; CP: coverage probability of $95\%$ pointwise confidence interval.}
\end{table}

\begin{table}[p]
\caption{Pointwise simulation results for the estimated functional regression parameters $\hat{\beta}_{n0,2,1}(\cdot)$ based on the proposed approach and the approach by \citet{fine2004temporal} (FYK04) for independent observations. Simulation scenario with a more variable cluster size $M\in\{5,\ldots,105\}$.}
\label{t:suppfour}
\begin{center}
\begin{tabular}{llcccccccc}
\hline
& & \multicolumn{4}{c}{Proposed} & \multicolumn{4}{c}{FYK04}\\
$n$ & $t$ & Bias & ASE & MCSD & CP & Bias &  ASE & MCSD & CP\\ 
\hline
50	&0.2	&-0.008	&0.054	&0.053	&0.956	&-0.037	&0.035	&0.050	&0.714\\
	&0.4	&-0.009	&0.058	&0.060	&0.934	&-0.042	&0.038	&0.054	&0.719\\
	&0.6	&-0.012	&0.061	&0.063	&0.927	&-0.045	&0.041	&0.057	&0.727\\
	&0.8	&-0.012	&0.064	&0.068	&0.915	&-0.046	&0.043	&0.060	&0.739\\
	&1.0	&-0.014	&0.067	&0.068	&0.942	&-0.048	&0.046	&0.060	&0.763\\
	&1.2	&-0.013	&0.070	&0.073	&0.941	&-0.050	&0.048	&0.063	&0.778\\
	&1.4	&-0.012	&0.072	&0.078	&0.925	&-0.048	&0.050	&0.067	&0.781\\
	&1.6	&-0.011	&0.074	&0.078	&0.928	&-0.050	&0.053	&0.069	&0.766\\
	&1.8	&-0.008	&0.077	&0.082	&0.932	&-0.047	&0.055	&0.071	&0.802\\
100	&0.2	&-0.002	&0.039	&0.039	&0.939	&-0.034	&0.024	&0.036	&0.639\\
	&0.4	&-0.002	&0.041	&0.043	&0.942	&-0.037	&0.027	&0.038	&0.639\\
	&0.6	&-0.003	&0.044	&0.046	&0.938	&-0.040	&0.028	&0.041	&0.645\\
	&0.8	&-0.005	&0.046	&0.048	&0.949	&-0.043	&0.030	&0.042	&0.667\\
	&1.0	&-0.005	&0.048	&0.049	&0.939	&-0.044	&0.032	&0.043	&0.671\\
	&1.2	&-0.003	&0.050	&0.051	&0.945	&-0.043	&0.034	&0.045	&0.708\\
	&1.4	&-0.004	&0.052	&0.054	&0.933	&-0.043	&0.035	&0.047	&0.713\\
	&1.6	&-0.004	&0.054	&0.055	&0.940	&-0.044	&0.037	&0.047	&0.724\\
	&1.8	&-0.005	&0.055	&0.057	&0.941	&-0.045	&0.038	&0.050	&0.727\\
200	&0.2	&-0.001	&0.028	&0.028	&0.941	&-0.034	&0.017	&0.026	&0.500\\
	&0.4	&0.000	&0.029	&0.030	&0.940	&-0.036	&0.019	&0.026	&0.522\\
	&0.6	&-0.001	&0.031	&0.031	&0.942	&-0.038	&0.020	&0.028	&0.511\\
	&0.8	&0.000	&0.033	&0.034	&0.932	&-0.039	&0.021	&0.029	&0.544\\
	&1.0	&-0.001	&0.034	&0.035	&0.949	&-0.040	&0.023	&0.030	&0.570\\
	&1.2	&-0.001	&0.035	&0.035	&0.952	&-0.041	&0.024	&0.031	&0.560\\
	&1.4	&0.000	&0.037	&0.037	&0.948	&-0.042	&0.025	&0.032	&0.583\\
	&1.6	&-0.001	&0.038	&0.040	&0.932	&-0.042	&0.026	&0.034	&0.609\\
	&1.8	&-0.001	&0.039	&0.042	&0.933	&-0.043	&0.027	&0.035	&0.606\\
400	&0.2	&-0.001	&0.019	&0.020	&0.952	&-0.032	&0.012	&0.018	&0.315\\
	&0.4	&-0.002	&0.021	&0.021	&0.955	&-0.037	&0.013	&0.018	&0.282\\
	&0.6	&-0.001	&0.022	&0.022	&0.951	&-0.039	&0.014	&0.019	&0.276\\
	&0.8	&-0.002	&0.023	&0.024	&0.931	&-0.041	&0.015	&0.020	&0.281\\
	&1.0	&-0.002	&0.024	&0.025	&0.943	&-0.042	&0.016	&0.022	&0.305\\
	&1.2	&-0.002	&0.025	&0.026	&0.945	&-0.042	&0.017	&0.022	&0.332\\
	&1.4	&-0.002	&0.026	&0.026	&0.953	&-0.043	&0.017	&0.022	&0.340\\
	&1.6	&-0.003	&0.027	&0.027	&0.943	&-0.044	&0.018	&0.023	&0.359\\
	&1.8	&-0.001	&0.028	&0.028	&0.940	&-0.043	&0.019	&0.024	&0.405\\
\hline
\end{tabular}
\end{center}
{\footnotesize Note: $n$: number of clusters with cluster size $M\in\{5,\ldots,105\}$; ASE: average estimated standard error; MCSD: Monte Carlo standard deviation of the estimates; CP: coverage probability of $95\%$ pointwise confidence interval.}
\end{table}

\begin{table}[p]
\caption{Pointwise simulation results for the estimated functional regression parameters $\hat{\beta}_{n,2,2}(\cdot)$ based on the proposed approach and the approach by \citet{fine2004temporal} (FYK04) for independent observations. Simulation scenario with a more variable cluster size $M\in\{5,\ldots,105\}$.}
\label{t:suppfive}
\begin{center}
\begin{tabular}{llcccccccc}
\hline
& & \multicolumn{4}{c}{Proposed} & \multicolumn{4}{c}{FYK04}\\
$n$ & $t$ & Bias & ASE & MCSD & CP & Bias &  ASE & MCSD & CP\\ 
\hline
50	&0.2	&-0.005	&0.047	&0.047	&0.941	&-0.036	&0.026	&0.046	&0.599\\
	&0.4	&-0.006	&0.049	&0.052	&0.927	&-0.040	&0.029	&0.050	&0.608\\
	&0.6	&-0.006	&0.052	&0.054	&0.931	&-0.042	&0.031	&0.052	&0.626\\
	&0.8	&-0.008	&0.054	&0.058	&0.924	&-0.043	&0.033	&0.054	&0.633\\
	&1.0	&-0.007	&0.056	&0.058	&0.938	&-0.045	&0.035	&0.054	&0.656\\
	&1.2	&-0.009	&0.059	&0.062	&0.918	&-0.046	&0.037	&0.056	&0.662\\
	&1.4	&-0.007	&0.060	&0.065	&0.920	&-0.045	&0.038	&0.059	&0.675\\
	&1.6	&-0.007	&0.062	&0.064	&0.935	&-0.047	&0.040	&0.059	&0.703\\
	&1.8	&-0.008	&0.064	&0.067	&0.926	&-0.047	&0.042	&0.062	&0.699\\
100	&0.2	&-0.003	&0.033	&0.035	&0.937	&-0.034	&0.018	&0.033	&0.492\\
	&0.4	&-0.003	&0.035	&0.036	&0.940	&-0.037	&0.020	&0.034	&0.510\\
	&0.6	&-0.003	&0.037	&0.040	&0.935	&-0.039	&0.022	&0.037	&0.542\\
	&0.8	&-0.003	&0.039	&0.042	&0.918	&-0.041	&0.023	&0.038	&0.559\\
	&1.0	&-0.004	&0.040	&0.042	&0.932	&-0.042	&0.024	&0.039	&0.562\\
	&1.2	&-0.006	&0.042	&0.045	&0.927	&-0.044	&0.026	&0.040	&0.557\\
	&1.4	&-0.005	&0.043	&0.046	&0.939	&-0.044	&0.027	&0.040	&0.596\\
	&1.6	&-0.004	&0.044	&0.046	&0.939	&-0.044	&0.028	&0.041	&0.618\\
	&1.8	&-0.005	&0.046	&0.048	&0.938	&-0.045	&0.029	&0.042	&0.621\\
200	&0.2	&-0.001	&0.024	&0.024	&0.950	&-0.033	&0.013	&0.023	&0.371\\
	&0.4	&-0.002	&0.025	&0.025	&0.947	&-0.037	&0.014	&0.024	&0.353\\
	&0.6	& 0.000	&0.026	&0.027	&0.947	&-0.038	&0.015	&0.025	&0.360\\
	&0.8	&-0.001	&0.028	&0.029	&0.944	&-0.040	&0.016	&0.026	&0.377\\
	&1.0	&-0.001	&0.029	&0.029	&0.949	&-0.041	&0.017	&0.026	&0.410\\
	&1.2	&-0.001	&0.030	&0.030	&0.944	&-0.042	&0.018	&0.027	&0.414\\
	&1.4	&-0.001	&0.031	&0.032	&0.944	&-0.042	&0.019	&0.028	&0.428\\
	&1.6	& 0.000	&0.032	&0.033	&0.949	&-0.042	&0.020	&0.029	&0.461\\
	&1.8	&-0.001	&0.033	&0.034	&0.943	&-0.043	&0.021	&0.029	&0.457\\
400	&0.2	&-0.001	&0.017	&0.017	&0.948	&-0.032	&0.009	&0.016	&0.195\\
	&0.4	&-0.001	&0.018	&0.018	&0.954	&-0.036	&0.010	&0.017	&0.158\\
	&0.6	&-0.002	&0.019	&0.018	&0.950	&-0.039	&0.011	&0.017	&0.137\\
	&0.8	&-0.001	&0.020	&0.020	&0.945	&-0.040	&0.011	&0.018	&0.161\\
	&1.0	&-0.002	&0.021	&0.021	&0.944	&-0.042	&0.012	&0.018	&0.165\\
	&1.2	&-0.001	&0.021	&0.021	&0.954	&-0.042	&0.013	&0.019	&0.188\\
	&1.4	&-0.001	&0.022	&0.022	&0.953	&-0.043	&0.013	&0.020	&0.199\\
	&1.6	&-0.001	&0.023	&0.023	&0.946	&-0.043	&0.014	&0.020	&0.204\\
	&1.8	&-0.001	&0.023	&0.023	&0.944	&-0.043	&0.015	&0.020	&0.225\\
\hline
\end{tabular}
\end{center}
{\footnotesize Note: $n$: number of clusters with cluster size $M\in\{5,\ldots,105\}$; ASE: average estimated standard error; MCSD: Monte Carlo standard deviation of the estimates; CP: coverage probability of $95\%$ pointwise confidence interval.}
\end{table}

\begin{table}[p]
\caption{Simulation results on the coverage probabilities of the $95\%$ simultaneous confidence bands for the proposed method and the method by \citet{fine2004temporal} (FYK04) for independent observations. Simulation scenario with a more variable cluster size $M\in\{5,\ldots,105\}$.}
\label{t:suppsix}
\begin{center}
\begin{tabular}{lcccccc}
\hline
& \multicolumn{3}{c}{Proposed} & \multicolumn{3}{c}{FYK04} \\
 $n$ & $\beta_{0,2,0}(t)$ & $\beta_{0,2,1}(t)$ & $\beta_{0,2,2}(t)$ & $\beta_{0,2,0}(t)$ & $\beta_{0,2,1}(t)$ & $\beta_{0,2,2}(t)$ \\
\hline
\\[0.1ex]
50	&0.921	&0.905	&0.907	&0.051	&0.648	&0.527\\
100	&0.922	&0.934	&0.918	&0.007	&0.560	&0.428\\
200	&0.942	&0.929	&0.942	&0.000	&0.432	&0.271\\
400	&0.945	&0.943	&0.948	&0.000	&0.184	&0.081\\
\hline\\
\end{tabular}
\end{center}
{\footnotesize Note: $n$: number of clusters with the cluster size $M\in\{5,\ldots,105\}$.}
\end{table}

\begin{table}[p]
\caption{Simulation results on the empirical type I error rates $(H_0)$ and the power levels $(H_1)$ for the proposed Kolmogorov--Smirnov-test and the corresponding test proposed by \citet{fine2004temporal} (FYK04) for independent observations, at the $\alpha=0.05$ level. Simulation scenario with a more variable cluster size $M\in\{5,\ldots,105\}$.}
\label{t:suppseven}
\begin{center}
\begin{tabular}{lcccc}
\hline
 & \multicolumn{2}{c}{Proposed} & \multicolumn{2}{c}{FYK04} \\[1ex]
$n$ & $\beta_{0,2,1}(t)$ & $\beta_{0,2,2}(t)$ & $\beta_{0,2,1}(t)$ & $\beta_{0,2,2}(t)$\\[1ex]
\hline
 \multicolumn{5}{c}{$H_{0}: \beta_{0,2,l}(t)=0,\quad l=1,2$} \\[1ex]
50	&0.091	&0.082	&0.182	&0.263\\
100	&0.078	&0.077	&0.223	&0.349\\
200	&0.074	&0.055	&0.293	&0.434\\
400	&0.051	&0.048	&0.459	&0.646\\
\hline
\multicolumn{5}{c}{$H_{1}: \beta_{0,2,l}(t)=-0.05,\quad l=1,2$}\\[1ex]
50	&0.313	&0.381	&0.656	&0.835\\
100	&0.434	&0.581	&0.894	&0.972\\
200	&0.639	&0.826	&0.986	&0.999\\
400	&0.936	&0.994	&1.000	&1.000\\
\hline
\multicolumn{5}{c}{$H_{1}: \beta_{0,2,l}(t)=-0.1,\quad l=1,2$}\\[1ex]
50	&0.694	&0.806	&0.957	&0.993\\
100	&0.890	&0.974	&0.999	&1.000\\
200	&0.993	&1.000	&1.000	&1.000\\
400	&1.000	&1.000	&1.000	&1.000\\
\hline
\multicolumn{5}{c}{$H_{1}: \beta_{0,2,l}(t)=-0.25,\quad l=1,2$}\\[1ex]
50	&1.000	&1.000	&1.000	&1.000\\
100	&1.000	&1.000	&1.000	&1.000\\
200	&1.000	&1.000	&1.000	&1.000\\
400	&1.000	&1.000	&1.000	&1.000\\
\hline\\
\end{tabular}
\end{center}
{\footnotesize Note: $n$: number of clusters with the cluster size $M\in\{5,\ldots,105\}$.}
\end{table}

\subsection{Additional SPECTRUM Trial Data Analysis Results}

\begin{figure}[p]
\begin{center}
 \includegraphics[width=0.8\textwidth]{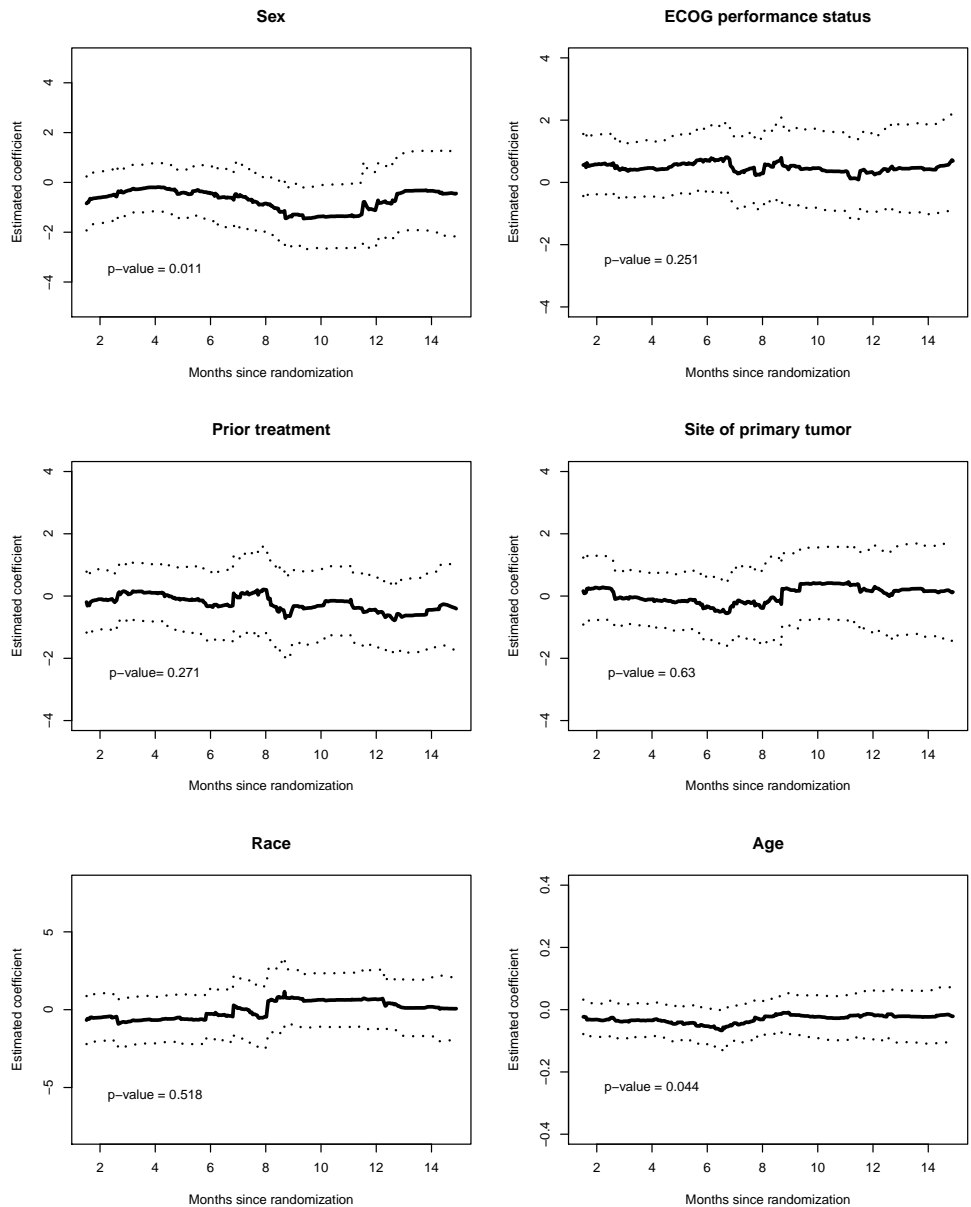}
\caption{Plots for the estimated regression coefficients of sex, ECOG performance status, prior treatment history, site of primary tumor, race, and age, for the "tumor response" state, along with the corresponding 95\% simultaneous confidence bands and p-values based the proposed methodology.}
\label{f:three}
\end{center}
\end{figure}

\begin{figure}[p]
\begin{center}
 \includegraphics[width=0.8\textwidth]{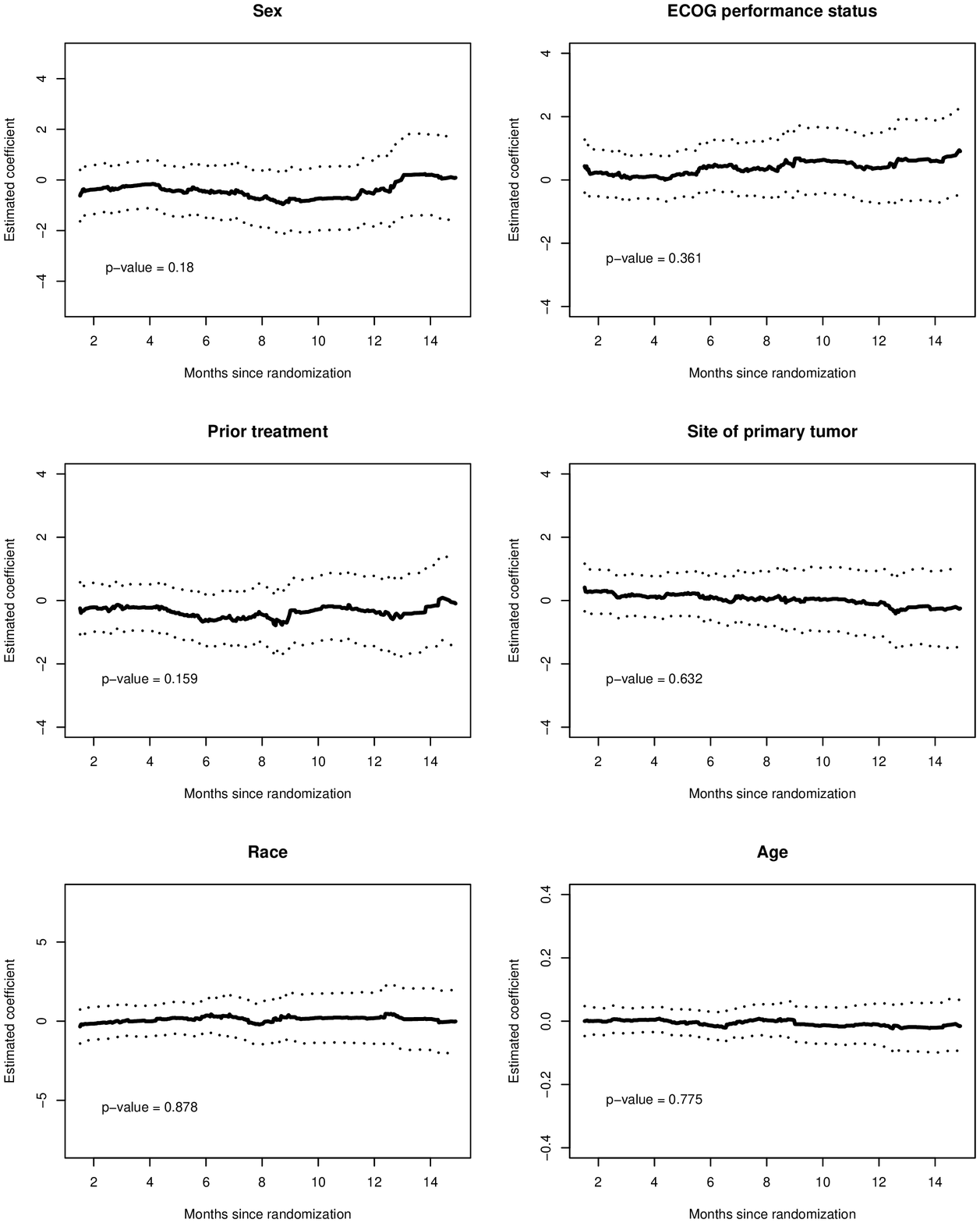}
\caption{Plots for the estimated regression coefficients of sex, ECOG performance status, prior treatment history, site of primary tumor, race, and age, for the "tumor response" state, along with the corresponding 95\% simultaneous confidence bands and p-values based on methodology by \citet{fine2004temporal} for independent observations.}
\label{f:four}
\end{center}
\end{figure}

Plots of the estimated time-varying regression coefficients of sex, ECOG performance status, prior treatment history, site of primary tumor, race, and age, for the "tumor response" state, along with the corresponding 95\% simultaneous confidence bands and p-values based on the proposed methodology are presented in Figure~\ref{f:three}. For comparison, the corresponding plots based on the previous method by \citet{fine2004temporal} for independent observations are provided in Figure~\ref{f:four}.

\end{document}